# Revealing Polymorph-Specific Transduction in $WO_3$ during Acetone Sensing

Matteo D'Andria[†,1], Meng Yin[†,2], Stefan Neuhauser[1], Vlasis G. Mavrantzas[3,4], Ying Chen[5], Ken Suzuki[6], and Andreas T. Güntner[1,*]

[1] *Human-centered Sensing Laboratory, Department of Mechanical and Process Engineering, ETH Zurich, CH-8092 Zurich, Switzerland*
[2] *Department of Finemechanics, Graduate School of Engineering, Tohoku University, Sendai, Miyagi 9808579, Japan*
[3] *Particle Technology Laboratory, Department of Mechanical and Process Engineering, ETH Zurich, CH-8092 Zurich, Switzerland*
[4] *Department of Chemical Engineering, University of Patras, Patras, GR 26504, Greece*
[5] *Global Learning Center, Tohoku University, Sendai, Miyagi 9808576, Japan*
[6] *Green X-tech Center, Green Goals Initiative, Tohoku University, Sendai, Miyagi 9808579, Japan*

[†] These authors contributed equally to this work as co-first authors.

*corresponding author: andregue@ethz.ch

# Abstract

Polymorphs are distinct structural forms of the same compound and offer unique opportunities to tailor material properties without altering chemical composition. In particular, the polymorphs of $WO_3$ have been widely explored for their molecular sensing performance; yet, the mechanistic aspects behind their different chemoresistive properties have remained elusive or poorly understood. Here, we highlight the energetic allocation of transferred charge as a critical aspect for chemoresistive response generation, providing a new perspective beyond more conventional net-transfer metrics, which are usually deployed to investigate gas-solid interactions. To this, we combined *operando* work function, chemisorption analysis, and *in situ* spectroscopy with density functional theory calculations on the example of acetone. Both $\gamma$- and $\varepsilon$-$WO_3$ exhibit comparable surface-level activation of acetone, mediated by electron-deficient, coordinatively unsaturated tungsten sites. However, only $\varepsilon$-$WO_3$ stabilizes analyte-induced electronic states derived from W(5d) orbitals lying just below the conduction band — an energetically favourable region for conductivity modulation under operating conditions. While being associated with marginal work function shifts, these states reflect deeper subsurface electronic rearrangements that may underlie the $\varepsilon$-$WO_3$'s superior transduction efficiency despite similar receptor chemistry. Our results offer a new framework for rational transducer development rooted in intrinsic electronic structure.

# 1  Introduction

Thermally activated transformations on nanostructured surfaces, exploited in chemoresistive gas sensing,[1] heterogeneous catalysis[2] or energy conversion,[3] are fundamentally governed by charge transfer and redistribution.[4] These electron-driven processes underpin all oxidation chemistry[5] — from model reactions such as the total oxidation of CO[6] and larger volatile organics (e.g., alcohols,[7] ketones and aromatics[8]), to more challenging selective oxidations, including the conversion of methanol to formaldehyde,[9] and ammonia to nitrogen[10] or nitrous oxide.[11] Furthermore, electron transfer governs the binding strength of reactants,[12] the relative stability of intermediates,[13] and ultimately product distribution under reaction conditions.[14]

Mechanistic studies usually emphasize as performance descriptors at the gas-solid interface reactant conversion, product selectivity, and changes in oxidation state, but often ignore the role of electron transfer in modifying the electronic structure of semiconductor metal oxides ($MO_x$). For supported metal catalysts, electronic descriptors such as $d$-band center ($\varepsilon_d$) relative to the Fermi level ($E_F$) have been widely used to rationalize adsorbate energetics[15,16] and catalytic trends.[17] However, this description is limited in $d^0$-type[18] $MO_x$ such as $WO_3$,[19] where $\varepsilon_d$ is either ill-defined or irrelevant, due to the absence of a pseudo-continuum of occupied $d$-states near $E_F$. Furthermore, the formation of localized electronic states[20] induced by charge transfer within the adsorbate-substrate complex is rarely resolved with respect to intrinsic band structure quantities,[21] including the valence and conduction band edges or mid-gap states.

In heterogeneous catalysis, resolving how transferred electrons are accommodated within the material is often secondary to understanding the dynamics of surface-bound intermediates and selective bond activation.[22] In contrast, in chemoresistive sensors based on semiconductors, the key observable is the change in electrical resistance[23] of the solid itself, which directly reflects modifications in the electronic structure upon analyte-surface interaction. In this context, charge transfer not only affects surface chemistry but may also induce the formation of localized states within the band structure, thereby altering conductivity.

Tungsten trioxide ($WO_3$) is among the most studied $MO_x$ for chemoresistive gas (e.g., acetone,[24-27] $NO_2$,[28] isoprene,[29] toluene[30]) detection, with numerous reports[31-34] discussing its thermodynamically stable $\gamma$-phase. Also studied is the metastable $\varepsilon$-$WO_3$ polymorph[35] that has consistently shown higher responses, for instance to acetone, under comparable conditions.[36,37] This crystal phase has been implemented in advanced sensing platforms, including clinical tests for non-invasive metabolic monitoring *via* skin- and breath-acetone analysis[38-40] and has been deployed in diverse contexts, such as locating entrapped individuals[41] and supporting patient care in clinical settings,[42] with potential for future use in monitoring astronauts' health during space missions.[43] Already in its original synthesis report as a molecular sensor,[44] enhanced performance of $\varepsilon$-$WO_3$ was tentatively linked to ferroelectric surface domains, suggesting a coupling between crystal structure and sensing function. Yet, and despite considerable progress, the mechanistic origin of $\varepsilon$-$WO_3$'s superior response remains still unclear. Conventional surface-chemical[45] insights — based on adsorbate binding, oxygen activation, or intermediate speciation — have not captured the performance gap, while the impact of polymorphism on electron transfer and localized electronic states remains unexplored.

Here, we investigate how structural polymorphism governs the sensing response of $WO_3$ by an in-depth comparison of its $\gamma$- and $\varepsilon$-phases on the example of acetone. Through a combination of *in situ* spectroscopy, chemisorption analysis, electronic structure calculations, and simultaneous electrophysical measurements (work function, $\varphi$, and DC resistance, $R$), we relate analyte interaction to changes in the material's electronic structure. This enables us to move beyond conventional surface affinity considerations (adsorption energies) by assessing not only the extent of charge transfer (Bader charge), but also how and where charge is accommodated within the band structure to constitute a chemoresistive sensing response.

## 2   Results and Discussion

### 2.1   Material structural identity of WO$_3$ polymorphs

We harnessed non-equilibrium combustion-aerosol technology in the form of flame spray pyrolysis (FSP),[46] where rapid formation and quenching of nanoscale oxides kinetically trap $\varepsilon$-WO$_3$,[44] as similarly shown also for pseudo-binary metal oxides.[47] $\varepsilon$-WO$_3$ is the low-temperature (< – 50 °C) monoclinic phase of WO$_3$,[48] and is therefore metastable under ambient or sensing conditions. To ensure stability at operating temperatures, foreign element doping (i.e., Cr[44] or Si[36]) prevents restructuring into the thermodynamically stable $\gamma$-polymorph upon annealing. **Figure 1**a,b show the X-ray diffraction (XRD) patterns of WO$_3$ powders as synthesized without (a) and with (b) 10 mol% Si, along with structural refinement results. In agreement with the literature,[36] phase-pure $\gamma$- and $\varepsilon$-polymorphs are obtained, as confirmed by statistical fit quality ($R_\mathrm{p}$) and lattice parameters matching reference values (Table S1) with, however, possible $\varepsilon$-remnants in the pure WO$_3$ sample (Figure S1).

As a result, defect incorporation stabilizes phase-pure $\varepsilon$-WO$_3$ nanoparticles with slightly suppressed growth and larger surface areas[49] (Figure S2), in agreement with high-resolution transmission electron microscopy (HR-TEM) images in Figure 1c,d, both showing faceted nanoparticles and extended fringe patterns. The inset of Figure 1c indicates the $\gamma$-WO$_3$'s lattice planes (200) and (020) with interlayer spacings (*d*) of 3.7 Å and 3.8 Å, respectively. Similarly, the inset of Figure 1d highlights a nanocrystal with two sets of perpendicular fringes, as confirmed by its approximately cubic fast Fourier transform (FFT). This is consistent with the (002) and (110) spacings of an $\varepsilon$-WO$_3$ crystal oriented along [–110], where the fringe-specific assignment is tentative due to very similar *d*-values of ≈ 3.8 Å that are indistinguishable within the measuring accuracy of the method.

Given the large discrepancy in ionic radii[50] and charge imbalance between $W^{6+}$ and $Si^{4+}$, as well the differing coordination preference – octahedral[51] ($O_h$) $WO_6$ vs. tetrahedral[52] ($T_d$) $SiO_4$ – Si-related units are likely incorporated interstitially, forming non-substitutional local environments between $WO_6$ octahedra. This is supported by structural relaxation *via* density functional theory (DFT) calculations as shown in Figure 1e, where both *γ*- and *ε*-$WO_3$ frameworks exhibit tilted, corner-sharing $WO_6$. Silicon atoms in the *ε*-$WO_3$ structure are found outside the first tungsten shell, forming an incomplete $T_d$, distorted pyramid-like coordination with a missing bond, similarly obtained at different Si% (Figure S3). Such low-coordinated $SiO_x$ could induce local lattice distortions that kinetically stabilize *ε*-$WO_3$ under microstrained (0.253% deformation, Table S1) conditions.

To probe the tungsten coordination environments, we conducted $L_3$-edge X-ray absorption fine structure (XAFS) spectroscopy, extracting the Fourier-transformed (FT) $k^2$-weighted $\chi(k)$ signal, as shown in Figure 1f. Quantitative analysis of the first shell is summarized in Table S2. The two phases exhibit nearly identical W-O bond lengths and mean-square displacements, with coordination numbers of 5.48 and 5.24 for *γ*- and *ε*-$WO_3$, respectively, both lower than expected for ideal $O_h$ symmetry, likely due to oxygen vacancies ($V_O$) introduced during flame synthesis.[53]

The wavelet transform (WT) of *γ*-$WO_3$ (Figure 1g) features a broad *k*-distribution for 1.0 Å < R < 1.6 Å consistent with local $WO_6$ structural heterogeneity,[54] captured by modeling distinct crystallographic W-O environments[55] (Table S2). The WT of *ε*-$WO_3$ (Figure S4) similarly reveals inhomogeneous $WO_6$ coordination, accompanied by multiple scattering contributions for *R* > 1.7 Å. Subtracting the *γ*-$WO_3$'s contour map from that of *ε*-$WO_3$ (Figure S5a) highlights additional spectral intensity in the range 1.5 Å < *R* < 2.5 Å and at larger wavenumbers (> 9 Å$^{-1}$). These features are consistent with the presence of strained, $T_d$-like $SiO_x$ located beyond the first W coordination shell, as indicated by the WT signatures of *ad hoc* constructed scattering paths[56] including single and multi-scattering from Si atoms (Figure

S5b-d). Second-derivative W L$_3$-edge XANES analysis in Figure 1h shows identical ligand field-induced 5d splitting[55] and W$^{6+}$ oxidation states for γ- and ε-WO$_3$ (Figure S6), confirming that Si incorporation leaves the local WO$_6$ structure unperturbed.

## 2.2 Electronic structure of WO$_3$ polymorphs

A reliable electronic structure model is essential for understanding charge transfer and localized states in WO$_3$ polymorphs; we therefore first establish a validated DFT-based description, grounded in experimental observations. For ε-WO$_3$, Si atoms were not included in the model, consistent with its phase purity (Figure 1b) and the unaffected position of t$_{2g}$ and e$_g$ states (Figure 1h). UV-Vis diffuse reflectance spectra of γ- and ε-WO$_3$ (**Figure 2**a) display a steep absorption increase near 400 nm, attributed to O(2p) → W(5d) ligand-to-metal charge transfer,[57] reflecting valence band (VB) to conduction band (CB) excitations. Both polymorphs exhibit sharp optical profiles, consistent with phase-pure materials, although the presence of weakly optically active mid-gap defects, such as V$_O$-induced shallow donors,[58] cannot be excluded. Tauc analysis (see inset in Figure 2a) yields optical band gaps ($E_g$) of 3.07 eV and 3.20 eV for γ- and ε-WO$_3$, respectively, and the linearity of $(F(R_\infty) \cdot h\nu)^2$ vs. photon energy confirms the direct[59] VB → CB character, in agreement with literature.[60]

Next, the electronic band structures of γ- and ε-WO$_3$ were calculated using DFT. To address the well-known $E_g$ underestimation of standard GGA functionals,[61] the DFT-1/2 approach proposed by Ferreira and co-workers[62,63] was employed, which has also been used in previous works to describe γ-WO$_3$.[26,30] This method introduces half-occupation corrections to effectively capture self-energy effects without the need for empirical parameters, unlike DFT+U, and has proven reliable for weakly correlated oxides, such as WO$_3$.[26,64] The calculated band structures in Figure 2 for (b) γ- and (c) ε-WO$_3$ reveal direct band gaps located at the Γ-point for both polymorphs. DFT-1/2 yields $E_g$ of 2.99 eV and 3.11 eV for γ- and ε-WO$_3$, respectively, closely matching the optical gaps extracted from the Tauc analysis (Figure 2a, inset). These results also align well with previously reported experimental[48,65] values for γ-WO$_3$, and show markedly improved accuracy compared to DFT calculations[66,67] using LDA or GGA-PBE functionals without U-type corrections (see Table S3). Note that the modelled structures are also correct crystallographic representations of the two phases, as calculated lattice constants of $\gamma$-WO$_3$ and $\varepsilon$-WO$_3$ (see Table S4) are in excellent agreement with our results (Table S1), as well as other experimental data[48,65,68] and previous theoretical studies.[69]

Surface slab models of $\gamma$- and $\varepsilon$-WO$_3$ were constructed exposing the (001) facets to provide realistic platforms for investigating analyte-surface interactions.[70,71] Each slab comprised six atomic layers separated by > 18 Å vacuum to eliminate periodic artifacts. To reflect the inherent nonstoichiometry of operating surfaces, as well as to ensure the presence of under-coordinated sites capable of engaging in adsorption and electron transfer,[72] surface and bulk V$_O$ were introduced (W$_{48}$O$_{142}$ supercells, hereafter denoted as WO$_{3-x}$), as shown in Figure 2d. This approach is consistent with our W-L$_3$-XAFS analysis (Figure 1f) and the literature, recognizing that fully stoichiometric surfaces offer limited reactivity toward adsorbates under catalytic or sensing conditions.[73] Introduction of V$_O$ results in the appearance of occupied states slightly below the CB minimum (Figure 2e,f) primarily derived from W 5d orbitals. Depending on V$_O$ location (bulk or surface), these states exhibit different degrees of spatial localization across the slab, as confirmed by partial charge density mapping (Figure S7). The lower formation energies of surface- compared to subsurface-V$_O$ (Table S5) suggest their energetically shallow nature. Such defect-induced states, acting as electron donors, are expected[45,74] to play a critical role in modulating charge transfer processes at the surface of WO$_3$.

## 2.3 Acetone adsorption and surface-chemical interactions

To elucidate analyte-surface interactions, we combined first-principles modeling with spectroscopic and thermochemical analysis. Adsorption geometries on $\gamma$- and $\varepsilon$-WO$_{3-x}$ surfaces were determined by DFT relaxation of acetone molecules placed near V$_O$ (**Figure 3**a). In both polymorphs, acetone preferentially coordinates via its carbonyl oxygen to under-coordinated tungsten centers adjacent to $V_O$ sites. Interestingly, the calculated adsorption energies ($E_{ads}$) are comparable for both polymorphs (i.e., − 1.39 eV and − 1.42 eV for $γ$- and $ε$-WO$_{3-x}$, respectively). This is consistent with comparable net charge transfer from acetone to the substrate ($ΔQ$), and adsorption-induced polarization leading to localized electron density on the carbonyl oxygen, as reflected by changes of local Bader charge ($Δq_i$, see Table S6). Note that slab-size analysis in Figure S8 confirms that acetone binding strength is insensitive to surface area, validating the robustness of adsorption energetics.

*In situ* diffuse reflectance infrared Fourier transform spectroscopy (DRIFTS) during acetone exposure tracks the temperature-dependent surface chemistry of $γ$- and $ε$-WO$_3$ (Figure 3b). Gas-phase acetone typically exhibits a single, sharp $ν$(C=O) near 1740 cm$^{-1}$. Upon adsorption onto oxide surfaces, this carbonyl vibration undergoes electronic perturbation, resulting in multiple distinct[45] absorptions at 1712, 1726, and 1737 cm$^{-1}$, reflecting interaction with surface tungsten centers. A weaker component emerging at 1688 cm$^{-1}$ suggests a fraction of more strongly activated acetone species, possibly associated with partial C=O bond weakening. This interpretation is supported by increased fluctuations in the C=O bond length during a 10 ps *ab initio* molecular dynamics (AIMD) trajectory of acetone adsorbed on $γ$-WO$_3$ (001) at 500 K (Supplementary Note S1, Figure S9, S10).

At temperatures above 300 °C, the carbonyl-associated bands diminish while new features appear at 1600 cm$^{-1}$ and 1510 cm$^{-1}$. The 1600 cm$^{-1}$ band is consistent with the $ν$(C=C) of mesityl oxide,[75] an intermediate of acetone condensation, while the 1510 cm$^{-1}$ feature is attributed to enolate-like species.[76] Additional absorptions at 1463 and 1365 cm$^{-1}$ are tentatively assigned to carboxylate-like[77] $ν_{asym}$(COO$^−$) and $ν_{sym}$(COO$^−$), respectively, which — together with a broad $δ$(CH$_3$) shoulder near[78] 1380 cm$^{-1}$ — are consistent with the incipient formation of formate and/or acetate species.[74] This evolution reflects a stepwise acetone activation path, progressing from molecular adsorption to partial C-C coupling and oxygenate formation. Most importantly, across the temperature range including their typical operating conditions during acetone sensing (i.e., 300 – 400 °C[24-27,31-34]), $\gamma$- and $\varepsilon$-WO$_3$ exhibit closely similar surface reactivity, with only minor variations in band intensity and development.

Figure 3c shows thermal desorption profiles of acetone-exposed $\gamma$- and $\varepsilon$-WO$_3$ under He, monitoring $m/z$ = 43 (acetone), 44 (CO$_2$), and 18 (H$_2$O). Both polymorphs exhibit a sharp $m/z$ = 43 peak at about 80 °C, attributed to the release of physisorbed acetone. A secondary shoulder follows at ~ 150 °C and 180 °C for $\gamma$- and $\varepsilon$-polymorph, respectively, indicating slightly stronger retention or slower desorption kinetics in the $\varepsilon$-phase. Above 200 °C, oxygenated fragments begin to evolve: CO$_2$ release above 300 °C reflects lattice oxygen involvement in a Mars–van Krevelen (MvK) mechanism, consistent with WO$_3$-based catalysts for acetone sensing[28] as well as the expected role of V$_O$.[79] H$_2$O evolution spans a broad range that could originate from dehydrogenation of surface-bound acetates[80] or, in the case of $\varepsilon$-WO$_3$, desorption from Si-bonded hydroxyls introduced during FSP synthesis. Overall, these trends align with the IR spectra shown in Figure 3b, where oxygenate and condensation products are observed within similar temperature ranges.

Extended surface-chemical analysis further highlights the similarity of acetone interaction with $\gamma$- and $\varepsilon$-WO$_3$. Our results in Figure 3d (see also Figure S11-14 and Supplementary Note S2 for detailed discussion) confirm the formation of reduced surface W$^{(6-\delta)+}$ species[81] upon acetone adsorption onto under-coordinated tungsten sites, which feature strong Lewis-acidic character, as identified by thermochemical desorption of basic probing molecules,[82] i.e., NH$_3$ (Figure 3e) and pyridine (Py, Figure S11). Temperature-programmed reductions (TPR) under different environments (H$_2$, CO, acetone, Figure S12, S13) evidence a measurable acetone-induced redox activity around 300 – 400 °C, that is, a notably higher reactivity than for H$_2$ and CO (> 500 °C), possibly rationalizing their characteristically weak responses observed in WO$_3$ chemoresistors. Finally, temperature-programmed (re-)oxidation (TPO) in O$_2$ (Figure 3f,g) indicates the dynamicity and reversibility of V$_O$ population, which is a central aspect to MvK oxidations and additionally corroborated by temperature-programmed desorption (TPD) of $CO_2$ (Figure S14). Summa summarum, our extensive surface-chemical characterization indicates the rather identical surface properties of $\gamma$- and $\varepsilon$-polymorphs; next, we assess their chemoresistive behavior.

## 2.4 Transducer function and electronic structure analysis

Despite similar acetone surface interaction (Figure 3), $\varepsilon$-$WO_3$ consistently exhibits higher responses to acetone, for instance, at 1000 ppb and 330 °C in air (**Figure 4**a). This enhancement is observed between 10 – 1000 ppb (Figure 4b and Figure S15), where both polymorphs feature linear log-log response characteristics. The slightly lower power law exponent of $\varepsilon$-$WO_3$ suggests differences in the underlying charge transport or modulation mechanism,[83] which cannot be rationalized by surface reactivity that governs the receptor (i.e., acetone adsorption, see section 2.3),[27] but rather indicate that structural polymorphism directly affects the transducer function.

Figures 4c and d indicate the plane-averaged differential charge density $\Delta\rho(z)$ derived from DFT along the slab depth ($z$) for $\gamma$- and $\varepsilon$-$WO_{3-x}$ (001), respectively, highlighting how electronic rearrangements extend into the material upon acetone adsorption. Both materials exhibit a net electron transfer from acetone to $WO_{3-x}$ following molecular binding — consistent with increased conductivity in n-type $WO_3$. Yet, only $\varepsilon$-$WO_{3-x}$ shows more pronounced subsurface electronic perturbations, several atomic layers beneath the surface, extending into the bulk-like region of the slab. This suggests that charge redistribution reaches beyond few topmost layers into interior domains where electronic transport occurs. This is further supported by Figure S16 and S17, which compare differential charge densities across both stoichiometric and reduced surfaces. While CO induces only limited perturbations — also in the presence of $V_O$ that generally amplifies[84] $\Delta\rho$ amplitude (e.g., Figure S17a,b vs. Figure S17c,d) — acetone adsorption on $\varepsilon$-$WO_{3-x}$ in Figure 4d exhibits the most pronounced subsurface modulation. This reflects a synergistic interplay of polymorphism, $V_O$ population, and analyte-specific binding that results in the highest chemoresistive response. Note, however, that the $\Delta\rho(z)$ amplitude does not explain, for instance, $\gamma$-WO$_3$'s larger response to acetone compared to CO (see Figure 4c and Figure S17c). In fact, in this case, the more conventional surface-chemical approach is conclusive already, and provides both: (i) improved analyte-induced lattice reducibility (Figure S12a vs. S13a), as well as (ii) more favorable DFT-$E_{ads}$ of – 1.39 eV for acetone, that is markedly lower than – 0.63 eV obtained for CO.

The electronic depth-modulation of $\varepsilon$-WO$_{3-x}$ may be amplified by its incipient ferroelectricity,[48] where surface polar distortions propagate electrostatic effects deeper into the lattice — potentially stabilizing lattice-coupled charge carriers (e.g., polarons[85]) and enhancing subsurface charge accommodation. Dipole moment calculations (Table S7) show that V$_O$'s induce significantly larger polarization in $\varepsilon$- compared to $\gamma$-WO$_{3-x}$, consistent with a step-like vacuum potential offset across the slab — indicative of an asymmetric electrostatic profile between top and bottom facets (Figure S18). This inherent asymmetric charge distribution may also enhance electrostatic interactions with polar analytes, such as acetone, compared to weakly polar species like CO, favoring stronger alignment and charge transfer across the polar $\varepsilon$-WO$_{3-x}$ surface. As a result, this deep $\Delta\rho(z)$ redistribution could enable carrier percolation through a wider crystallite network, effectively enhancing the transducer response. These polymorph-specific features are corroborated by charge density difference (CDD) maps, which reveal extended electron accumulation around lattice O$^{2-}$ deeper in the $\varepsilon$-WO$_{3-x}$ slab, accompanied by a corresponding depletion in the topmost atomic planes. Hence, while structural polymorphism does not impact the adsorption strength ($E_{ads}$, Figure 3a) or net charge transfer, it determines how deeply the injected charge is accommodated — ultimately modulating conductivity (i.e., chemoresistive response), in agreement with our experimental observation (Figure 4a).

To further assess the transduction properties of γ- and ε-WO$_3$, we performed simultaneous work function (φ) and DC resistance measurements.[86] The results of our *operando* φ analysis (Figure S19–24) are summarized in **Figure 5**a,b, showing the resistance vs. Δφ (referenced to its value in dry synthetic air) attained in 0 – 20 vol% O$_2$/N$_2$ (triangles) and 0 – 50 ppm acetone/air (squares) mixtures. Note that the O$_2$/N$_2$ data for γ-WO$_3$ (Figure S19) were not included in Figure 5a as these do not follow a monotonic trend, and the *R* vs. Δφ space is most frequently used[87] to visualize Arrhenius-like (that is, thermally activated) transducer characteristics. As discussed in the Supplementary Note S3 and Figure S25, S26, such deviation for γ-WO$_3$ is attributed to molecularly adsorbed O$_2$ species, O$_2$(ad),[88-93] opposed to (iono-) sorbed O$_\beta^{\alpha-}$.

Exposing γ-WO$_3$ (Figure 5a) to 0 – 50 ppm acetone yields *R* vs. φ points following an exponential trend with an inverse coefficient of 1.35 in Boltzmann's factor, consistent with electron-depletion-controlled transduction, as classically observed in SnO$_2$-based systems.[94] As shown in Figure 5b, ε-WO$_3$ exhibits a pronounced φ drop (~ 80 meV) at 150 ppb acetone. However, at higher acetone concentrations, φ shifts become marginal and approach the noise floor, whereas resistance continues to drop — resulting in a steep, scattered regression line which systematically departs from the O$_2$/N$_2$ transduction characteristic.

In polycrystalline films of n-type chemoresistive materials, resistance is usually directly modulated by back-to-back Schottky barriers formed at the grain boundaries between individual nanoparticles.[89] There, CB electrons are trapped at oxygen-related surface acceptor states, leading to a less conductive, i.e., electron-depleted, space charge layer due to the CB energy barrier height ($qV_S$). Upon accepting electrons from a reducing agent into delocalized CB states above, for instance, the CB minimum (CBM), upwards $E_F$-shifts (or, equivalently, downwards φ-shifts) are reflected in lower $qV_S$ and, therefore, lower resistances.

Under typical sensing conditions (i.e., high-oxygen backgrounds), Boltzmann's statistics are valid[86] and the resistance is pinned to φ through a simple relation: $R \sim exp\left(\frac{\phi}{m \cdot k_B T}\right)$, as observed for widely studied $SnO_2$ in Figure S27-30 (see also Supplementary Note S4) and in agreement with the literature.[95] Therein, $m$ is a fitting parameter and regarded as a measure of transduction efficiency, explaining,[91] for instance, the lower sensitivity of p-type (theoretical $m=2$) compared to n-type (theoretical $m=1$) $MO_x$. The value observed for our $\varepsilon$-$WO_3$ (Figure 5b, acetone/air mixtures) of 0.44 is far below the range expected for the $E_F$-pinned transduction regime, exhibiting a clear divergence in $R$-$\varphi$ characteristics that, to the best of our knowledge, is observed here for the first time. This is consistent with transduction enhancement from deep-layer charge accommodation (Figure 4d) rather than classical band bending, and such seamless $\varphi$-shifts in $\varepsilon$-$WO_3$ may originate from populating localized sub-CBM states that do not appreciably shift $E_F$,[96] being energetically decoupled from the CB edge.

To investigate this further, we evaluated the electronic band structures of $\gamma$- and $\varepsilon$-$WO_{3-x}$ before and after analyte introduction by DFT.[97] In $\gamma$-$WO_{3-x}$ (Figure 5c), no discernible change is observed in the vicinity of the CBM; the band edges remain unaltered, and no evident additional states emerge. In contrast, $\varepsilon$-$WO_{3-x}$ (Figure 5d) features the appearance of well-defined electronic states just below the CBM following acetone adsorption. These states, predominantly of W(5d)-character, are thermally accessible and could be populated under operating conditions[98] — providing an energetically favourable means to conductivity modulation.[99] This orbital-energy-resolved distinction aligns with our hypothesis that the energetic allocation of transferred electrons — e.g., their proximity to the CBM — can be critical for chemoresistive response generation and needs to be considered next to conventional "total-count" metrics such as net charge transfer. As a result, polymorphism in $WO_3$ modulates chemoresistive sensing behavior primarily by shaping the density and accessibility of conduction-relevant states.

# 3 Conclusion

This study delivers a mechanistic framework to investigate polymorph-specific electronic transduction in semiconductive nanoparticulate films during molecular gas-solid interactions. It is applied to reveal the origin of a long-standing observation in molecular sensing with $WO_3$ polymorphs — namely, the enhanced response of the metastable $\varepsilon$- over the $\gamma$-phase for acetone — whose origin had remained elusive despite widespread technological use. By combining *in situ* spectroscopy, thermochemical desorption, and *operando* work function measurements with first-principles electronic structure calculations, we show that $\gamma$- and $\varepsilon$-$WO_3$ activate acetone similarly at the surface level via coordinatively unsaturated, electron-deficient tungsten sites. The stronger chemoresistive response of $\varepsilon$-$WO_3$ is associated to stabilized W(5d)-derived electronic states just below the conduction band minimum upon analyte adsorption. These states are thermally accessible and conduction-relevant, offering an energetically favorable pathway for resistance modulation under operating conditions.

Together, these findings underscore that structural polymorphism does not dictate analyte affinity in $WO_3$, but rather tunes the energetic landscape of electron accommodation — shifting the focus from classical charge-transfer metrics to orbital-level-resolved transduction. This advance paves the way for rational transducer design strategies grounded in electronic structure criteria that is applicable and likely relevant also to other chemoresistive materials, closing an important gap in mechanistic modeling and understanding.

# 4 Methods

## 4.1 Nanoparticle production

Nanoparticles of $\gamma$-$WO_3$ and $\varepsilon$-$WO_3$ were prepared by FSP, with a reactor design detailed elsewhere.[100] The $\varepsilon$-$WO_3$ phase was stabilized by Si addition (10 mol%).[36] To prepare the precursor, we dissolved ammonium metatungstate hydrate ($\geq$ 85% $WO_3$ gravimetric basis,

Sigma Aldrich, Switzerland) and hexamethyldisiloxane (Sigma Aldrich, Switzerland) in a 1:1 (by volume) mixture of ethanol and diethylene glycol monobutyl ether, to achieve a total molarity (W + Si) of 0.2 mol L$^{-1}$. Thereafter, the precursor was fed through a capillary and dispersed by $O_2$ (pressure drop of 1.6 bar) to form a fine spray. The precursor and dispersion flow rates were 5 mL min$^{-1}$ and 5 L min$^{-1}$, respectively. The spray was ignited and sustained by a pilot flame of premixed $CH_4$ (1.2 L min$^{-1}$, Methane 2.5, PanGas, Switzerland) and $O_2$ (3.2 L min$^{-1}$, Pangas, Switzerland). Additionally, 5 L min$^{-1}$ $O_2$ sheath flow was supplied to shield the flame and ensure excess oxidant. Gas flows were regulated by mass flow controllers (Bronkhorst, Netherlands), while the precursor solution flow was supplied by a syringe pump. The nanoparticles were deposited onto water-cooled glass fiber filters (257 mm diameter, GF6, Hahnemühle Fineart, Germany) at a height above the burner of 55 cm aided by a vacuum pump (Seco SV 1025 C, Busch, Switzerland). The particles were carefully removed from the filter with a spatula and the obtained powders were sieved with a 250 μm stainless steel mesh. To fabricate sensors, particles were directly deposited from the aerosol[100] for 4 minutes at 20 cm height above the burner onto water-cooled $Al_2O_3$ substrates, provided with interdigitated ($d$ = 250 μm) Pt electrodes for resistance readout (electrode type #103, Electronic Design Center, Case Western University, USA). Both as-produced nanoparticles and sensors were annealed in air at 500 °C for 5 hours (CWF 1300, Carbolite Gero, Germany).

## 4.2 Material characterization

XRD patterns were measured with a Bruker D2 phaser (Bruker, USA) diffractometer equipped with a Cu anode, operated at 30 kV and 10 mA. The powder samples were loaded onto low-background silicon holders and uniformly spread with the aid of a droplet of IPA. XRD patterns were recorded in Bragg-Brentano geometry at 2θ(Cu Kα) between 20 – 70 degrees, with a step size of 0.020 degrees and a time per step of 15 seconds. Powder diffractograms were analyzed by Rietveld refinement as implemented in Topas 4.2 (Bruker) software, using the crystallographic information files of $\gamma$-WO$_3$ (ICSD 80056) and $\varepsilon$-WO$_3$ (ICSD 84139). Peak broadening due to crystallite size ($d_{XRD}$) and microstrain[101] was modeled by Lorentzian and Gaussian contributions of a pseudo-Voigt peak shape, respectively, while instrumental broadening was accounted for through the fundamental parameter approach.[102]

N$_2$-physisorption isotherms (at 77 K) of powders (0.150 g) were recorded on a Tristar II Plus (Micromeritics, Germany). The specific surface area (SSA) was determined according to Brunauer-Emmett-Teller (BET) theory, at relative pressures between 0 – 0.4. Prior to measurement, the samples were degassed for 1.5 hours at 120 °C under N$_2$ to remove water adsorbates.

UV-Vis DRS was carried out with a Cary 5000 UV-Vis-NIR spectrophotometer (Agilent, USA). The powders were mixed with BaSO$_4$ to achieve ~10 wt%, and loaded in a high-temperature cell (CaF$_2$ windows) mounted in a Praying Mantis diffuse reflectance accessory (both Harrick Sci., USA). For (reactive) transient analysis, the diffuse reflectance (R) of the sample was recorded at a fixed wavelength of 600 nm, using a spectral band width of 2 nm and an averaging time of 2 seconds. The temperature was set by a controller and measured by a K-type thermocouple. Initially, powders were pre-treated in pure O$_2$ at 450 °C for 90 minutes, and cooled down under Ar to 300 °C. Therein, acetone vapor was introduced by bubbling Ar through liquid acetone (Sigma Aldrich, ≥ 99.5%) at 23 °C for 50 minutes, and the reduction rate constant ($k_{red}$) was estimated from the initial slope of the absorbance vs. time trace. The (optical) band gap ($E_g$) was estimated by acquiring diffuse reflectance spectra at room temperature between 200 – 800 nm, using Tauc's method, i.e., equation (1):[103,104]

$$(\alpha \cdot h\nu)^{1/\gamma} = B \cdot (h\nu - E_g) \qquad (1)$$

where $\gamma=1/2$ for direct electronic transitions and the absorption coefficient ($\alpha$) is estimated with the Kubelka-Munk (KM) function defined as in equation (2):

$$F(R_\infty) = \frac{(1-R_\infty)^2}{2R_\infty} \qquad (2)$$

DRIFT spectroscopy was performed on a Vertex 70v spectrometer (Bruker, USA) equipped with a liquid-nitrogen-cooled mercury cadmium telluride (MCT) detector. Spectral acquisition was carried out in diffuse reflectance mode (Harrick Sci. accessory) between 1000 – 4000 cm$^{-1}$ at 2 cm$^{-1}$ resolution and averaging 300 scans per spectrum. Therein, $C_3H_6O$ (2 vol% in Ar, Pangas) was supplied by a rotameter with a needle valve (30 mL min$^{-1}$) at temperatures between 50 – 450 °C.

Temperature-programmed experiments (heating rate of 10 K min$^{-1}$) were carried out with an Autochem III chemisorption analyzer (Micromeritics, Germany) equipped with a vapor generator, a thermal conductivity detector (TCD) and connected to an *online* quadrupole mass spectrometer (Omnistar, Pfeiffer, Switzerland). About 70 mg of sample were loaded into a U-shaped quartz reactor and pretreated in He at 300 °C. TPR under CO (10 mol% in He), $C_3H_6O$ (1000 ppm in Ar) and $H_2$ (5 vol% in Ar, all Pangas) were recorded between 30 – 1000 °C, and were followed by TPO under 5 vol% $O_2$ in He (Pangas). TPD runs were performed in He after adsorbing $O_2$, $CO_2$ (10 vol% in Ar), $C_3H_6O$ and Py, the latter generated as a vapor at a reflux temperature of 45 °C.

For HR-TEM, the material was dispersed in ethanol and a few drops of the suspension were deposited onto a perforated carbon foil supported on a copper grid. After solvent evaporation, the grid was mounted on the single tilt holder of the microscope. TEM investigations were performed on a JEOL JEM F300 (GrandARM) with a cold field emission gun operated at 300 kV.

XAFS of the W $L_3$-edge (10.207 keV) was carried out at the MAX IV (Lund, Sweden), using a Si(111) double crystal monochromator and ion chambers to record $I_0$ and $I_t$ signals. Samples were directly pressed into 13-mm-diameter pellets after mixing with boron nitride to optimize the transmission measurement. The data were processed using the Demeter software package[105] (including Athena and Artemis). Athena was used for unit-$E_0$-normalization and background removal to extract the $k^2$-weighted $\chi(k)$ signal. Artemis was used to fit the EXAFS data in real space between 2.5 Å$^{-1}$ < $k$ < 13.0 Å$^{-1}$ and 1.0 Å < $R$ < 2.1 Å with the multiple $k_w$ method.[106] The amplitude reduction factor $S_0^2$ from the EXAFS analysis of a W foil was 0.868, which was used as a fixed parameter for EXAFS fitting. The coordination numbers and bond lengths were calculated based on the reported structures from the Inorganic Crystal Structure Database (ICSD) indicated in the text.

## 4.3  Chemoresistive characterization

The sensors were mounted onto MACOR® holders and placed in a PTFE-made chamber.[107] The sensing film was heated by applying a constant voltage to a meander-shaped Pt heater in the back of the substrate. The temperature was determined with a multimeter (2700, Keithley, USA) by using the same Pt heater as the resistance temperature detector. The chamber was connected to a gas mixing set-up. Hydrocarbon-free synthetic air (Pangas, $C_nH_m$ and $NO_x$ < 100 ppb) was used as a carrier gas and the analytes from certified gas standards were admixed by mass flow controllers (Bronkhorst, Netherlands) to obtain the desired gas mixture composition. The calibrated and certified gas standards (all Pangas, dry synthetic air as carrier) used were $C_2H_6O$ (15 ppm), CO (506 ppm) and $C_3H_6O$ (18 ppm and 150 ppm), while the total flow was set constant to 300 mL min$^{-1}$. The DC resistance of the sensing film was measured continuously between the interdigitated Pt electrodes with a Keithley 2700 multimeter and Keithley DMM7510 picoammeter for $\gamma$- and $\varepsilon$-$WO_3$, respectively. The chemoresistive response ($S$) was defined in equation (3) as the normalized resistance variation:

$$S = \frac{R_a - R_g}{R_g} \qquad (3)$$

where $R_a$ and $R_g$ are the resistances of the sensing film under clean air and gas exposure, respectively.

## 4.4 *Operando* work function analysis

*Operando* work function measurements were performed by the Kelvin oscillator method[108] (single point KP020, KP Technology, UK). The sensors were mounted onto the same MACOR® holders and placed in an aluminum chamber, with a 4-mm hole drilling to allow the Au tip (2-mm diameter) of the Kelvin probe to approach the sample at working distances of about 0.5 mm, which was enabled by the gradient function of the control software. The ohmic resistance was simultaneously monitored by applying a probing voltage between 0.5 – 10 V and measuring the induced current (2400, Keithley). Prior to that, IV sweeps were recorded at operational condition to ensure ohmic behavior. The aluminum chamber, a lead of the Pt heater, as well as a lead of the chemoresistive film were connected and equipotential with the Kelvin probe. To control the sample gas environment, the accessory was installed in a glovebox, gases were supplied by the same certified standards described above, with the addition of pure $N_2$ (purity 5.0, Pangas) and $H_2$ (47.9 ppm in dry synthetic air, Pangas), and mixed with mass flow controllers (Brooks instrument, USA). The total flow was kept at 300 mL min$^{-1}$.

## 4.5 DFT calculations

DFT calculations were carried out with the Vienna Ab initio Simulation Package (VASP).[109] To account for the interaction between ion cores and valence electrons, the projector augmented wave (PAW) method was employed.[110] Electron exchange-correlation interactions were computed using the generalized gradient approximation (GGA) with the Perdew-Burke-Ernzerhof (PBE) functional,[111] along with the Grimme D3 dispersion correction.[112] The cutoff energy for the plane wave expansion was set to 450 eV after convergence testing. For geometry optimization, the conjugate gradient algorithm was used to relax atomic positions until the total force on each ion was smaller than 0.02 eV Å$^{-1}$, while the convergence criterion for the electronic self-consistency cycle was set to 10$^{-6}$ eV. The electronic properties of the relaxed structure were calculated using the DFT-½ method, adding to only oxygen atoms as in the previous literature.[26] To facilitate a more accurate comparison between $WO_3$ polymorphs, the *k*-point mesh density was kept consistent rather than maintaining an identical number of *k*-points, owing to differences in unit cell dimensions. Consequently, a Monkhorst-Pack grid of 7×7×6 was utilized for the *γ*-phase, while a denser grid of 9×10×7 was employed for the *ε*-phase, ensuring uniform sampling of the Brillouin zone.

Based on the relaxed bulk structures, slab models of *γ*- and *ε*-$WO_3$ surfaces were extracted by cleaving along the (001) direction.[70,71] For *γ*-$WO_3$, a c(2×2) reconstruction of the (001) surface was adopted, which is the most commonly used model in *γ*-$WO_3$ simulations.[113] These slab models consist of six atomic layers, containing 48 W atoms and 144 O atoms, with the top and bottom layers containing each half the oxygen atoms. All six atomic layers were fully relaxed during structural optimization to effectively eliminate mid-gap states, thereby avoiding the need for hydrogen termination on the bottom layer and preventing artifacts caused by geometry distortions from fixed bottom layers.[19] To ensure comparability between phases, a similar 2×2 supercell model with six layers was constructed for *ε*-$WO_3$ (001), comprising also 48 W atoms and 144 O atoms. Besides, a vacuum thickness of ~18 Å was introduced to prevent inter-slab interactions given that the calculations were performed under periodic boundary conditions. Spin polarization and dipole moment corrections were consistently applied throughout the calculations. Bulk and surface $V_O$'s were introduced as described in the text (section 2.2), and their formation energies in both polymorphs were computed by equation (4):

$$E_{formation}(V_O) = E_{defective} - E_{pristine} + \frac{1}{2}E_{O_2} \quad (4)$$

Further, the adsorption energy per molecule was calculated through the following equation:

$$E_{ads} = E_{substrate+adsorbate} - (E_{substrate} + E_{adsorbate}) \quad (5)$$

AIMD was performed using the canonical (NVT) ensemble with the Nosé-Hoover thermostat ($T$ = 500 K). Acetone dynamics on the surface of $WO_{3-x}$ was simulated for 10 ps with a time step of 1 fs.

## Author contributions

M.D. and A.T.G. conceived the research and designed the experiments. M.D., M.Y., and A.T.G. coordinated the study. M.Y., V.G.M., Y.C., and K.S. developed the computational framework. M.D. and S.N. were primarily responsible for experimental data collection and analysis. M.Y. analyzed the structures obtained from DFT calculations. A.T.G. supervised the project and was responsible for funding acquisition. M.D. and M.Y. prepared the figures and wrote the original draft. All authors supported the revision of the manuscript and gave final approval.

## Acknowledgements

This study was financially supported by the Innosuisse (Innovation project 109.063 IP-LS), Swiss State Secretariat for Education, Research, and Innovation (SERI) under contract number MB22.00041 (ERC-STG-21 "HEALTHSENSE"), the Swiss National Science Foundation (BRIDGE Discovery grant #218650) and JSPS KAKENHI (Grant number 23KJ0196). M.Y., Y.C. and K.S. acknowledge the Center for Computational Materials Science (CCMS), Institute for Materials Research (IMR), and Supercomputer facility AOBA (Tohoku University, Japan). M.D. acknowledges the Electrochemical Energy Systems Laboratory at ETH Zurich (Prof. Dr. M. Lukatskaya) for providing the glovebox for the operando work function analysis, Mr. Peter Feusi and Mr. Tiago Elias Abi-Ramia Silva for assistance with the installation of the operando work function setup. The authors acknowledge Dr. Frank Krumeich from the Scientific Center for Optical and Electron Microscopy (ScopeM) of ETH Zurich for support with TEM. M.D. gratefully acknowledges Dr. Stuart Ansell from the Balder beamline in Lund (MAX IV) for the support during measurement time granted under proposal ID 25240825.# References

1   Bulemo, P. M., Kim, D.-H., Shin, H., Cho, H.-J., Koo, W.-T., Choi, S.-J., Park, C., Ahn, J., Güntner, A. T. & Penner, R. M., *Chemical reviews* **2025**.
2   Anastas, P. T. & Kirchhoff, M. M., *Accounts of chemical research* **2002**, *35*, 686-694.
3   Huang, J., Yuan, Y., Shao, Y. & Yan, Y., *Nature Reviews Materials* **2017**, *2*, 1-19.
4   Cuenya, B. R. & Behafarid, F., *Surface Science Reports* **2015**, *70*, 135-187.
5   Védrine, J. C., *ChemSusChem* **2019**, *12*, 577-588.
6   He, C., Cheng, J., Zhang, X., Douthwaite, M., Pattisson, S. & Hao, Z., *Chemical reviews* **2019**, *119*, 4471-4568.
7   Mallat, T. & Baiker, A., *Chemical reviews* **2004**, *104*, 3037-3058.
8   Scirè, S. & Liotta, L. F., *Applied Catalysis B: Environmental* **2012**, *125*, 222-246.
9   Kim, T. H., Ramachandra, B., Choi, J. S., Saidutta, M., Choo, K. Y., Song, S.-D. & Rhee, Y.-W., *Catalysis letters* **2004**, *98*, 161-165.
10  Chen, L., Guan, X., Wu, X., Asakura, H., Hopkinson, D. G., Allen, C., Callison, J., Dyson, P. J. & Wang, F. R., *Proceedings of the National Academy of Sciences* **2024**, *121*, e2404830121.
11  Tang, Z., Surin, I., Rasmussen, A., Krumeich, F., Kondratenko, E. V., Kondratenko, V. A. & Pérez-Ramírez, J., *Angewandte Chemie International Edition* **2022**, *61*, e202200772.
12  Nam, D.-H., De Luna, P., Rosas-Hernández, A., Thevenon, A., Li, F., Agapie, T., Peters, J. C., Shekhah, O., Eddaoudi, M. & Sargent, E. H., *Nature materials* **2020**, *19*, 266-276.
13  Zhang, S., Fan, Q., Xia, R. & Meyer, T. J., *Accounts of chemical research* **2020**, *53*, 255-264.
14  Koper, M. T., *Chemical science* **2013**, *4*, 2710-2723.
15  Nørskov, J. K., Abild-Pedersen, F., Studt, F. & Bligaard, T., *Proceedings of the National Academy of Sciences* **2011**, *108*, 937-943.
16  Hammer, B. & Nørskov, J. K. in *Advances in catalysis* Vol. 45   71-129 (Elsevier, 2000).
17  Greeley, J., Stephens, I., Bondarenko, A., Johansson, T. P., Hansen, H. A., Jaramillo, T., Rossmeisl, J., Chorkendorff, I. & Nørskov, J. K., *Nature chemistry* **2009**, *1*, 552-556.
18  Greiner, M. T., Helander, M. G., Tang, W.-M., Wang, Z.-B., Qiu, J. & Lu, Z.-H., *Nature materials* **2012**, *11*, 76-81.
19  Wang, F., Di Valentin, C. & Pacchioni, G., *The Journal of Physical Chemistry C* **2012**, *116*, 10672-10679.
20  Leonov, I. & Biermann, S., *Physical Review B* **2021**, *103*, 165108.
21  Zhong, Z. & Hansmann, P., *Physical Review X* **2017**, *7*, 011023.
22  Nørskov, J. K., Bligaard, T., Logadottir, A., Bahn, S., Hansen, L. B., Bollinger, M., Bengaard, H., Hammer, B., Sljivancanin, Z. & Mavrikakis, M., *Journal of catalysis* **2002**, *209*, 275-278.
23  Barsan, N., Koziej, D. & Weimar, U., *Sensors and Actuators B: Chemical* **2007**, *121*, 18-35.
24  Lei, G., Lou, C., Liu, X., Xie, J., Li, Z., Zheng, W. & Zhang, J., *Sensors and actuators B: Chemical* **2021**, *341*, 129996.
25  Schmitt, E. A., Krott, M., Epifani, M., Suematsu, K., Weimar, U. & Barsan, N., *The Journal of Physical Chemistry C* **2024**, *128*, 1633-1643.
26  Americo, S., Pargoletti, E., Soave, R., Cargnoni, F., Trioni, M. I., Chiarello, G. L., Cerrato, G. & Cappelletti, G., *Electrochimica Acta* **2021**, *371*, 137611.
27  Staerz, A., Weimar, U. & Barsan, N., *Sensors and Actuators B: Chemical* **2022**, *358*, 131531.
28  Staerz, A., Berthold, C., Russ, T., Wicker, S., Weimar, U. & Barsan, N., *Sensors and Actuators B: Chemical* **2016**, *237*, 54-58.
29  Weber, I. C., Oosthuizen, D. N., Mohammad, R. W., Mayhew, C. A., Pratsinis, S. E. & Güntner, A. T., *ACS sensors* **2023**, *8*, 2618-2626.
30  Trioni, M. I., Cargnoni, F., Americo, S., Pargoletti, E., Chiarello, G. L. & Cappelletti, G., *Nanomaterials* **2022**, *12*, 2696.
23

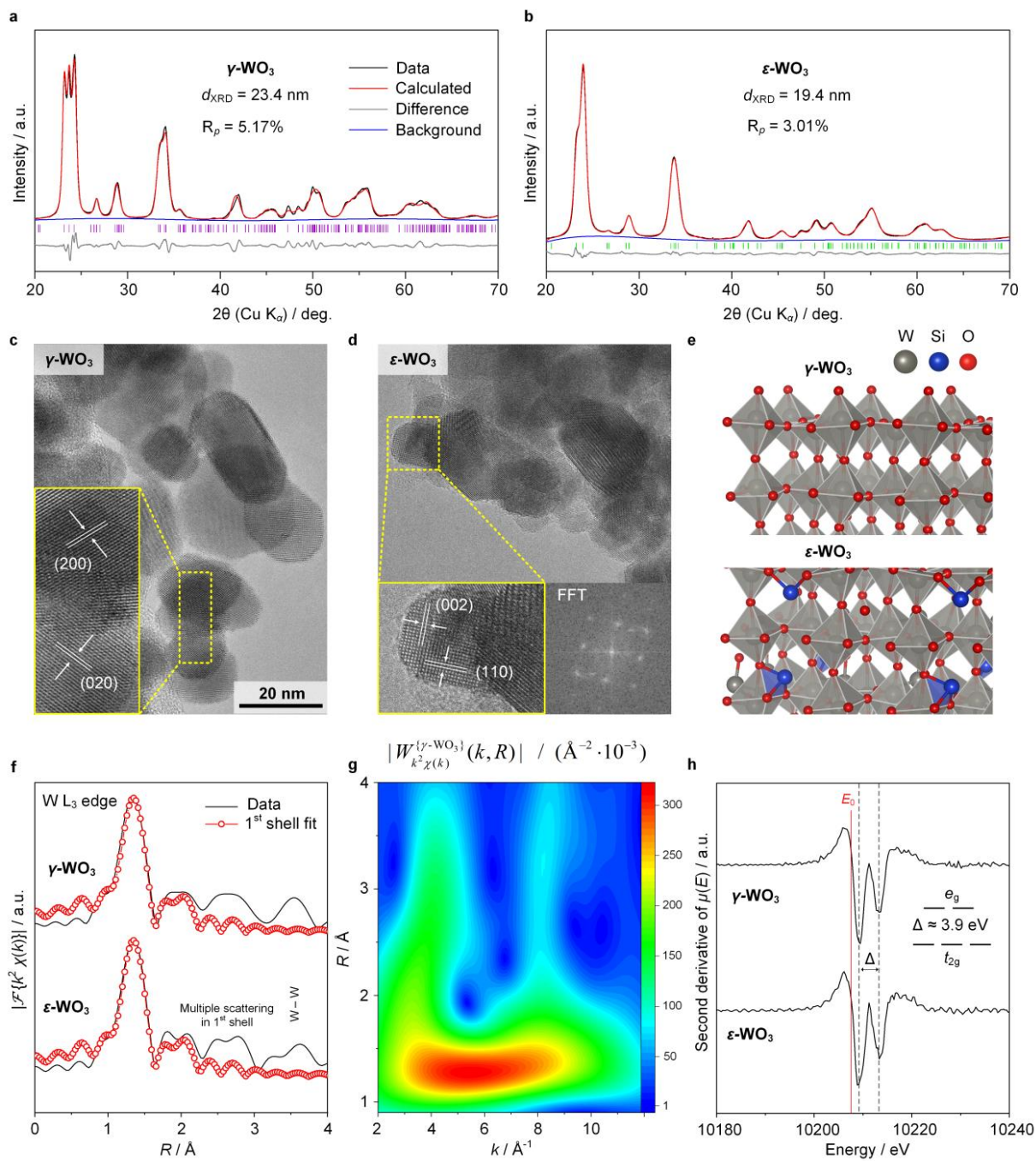

**Figure 1 | Material characterization of WO$_3$ polymorphs.** XRD patterns and Rietveld refinements for (a) $\gamma$-WO$_3$ as well as (b) $\varepsilon$-WO$_3$ with 10 mol% Si, with indicated coherent crystallite sizes of WO$_3$ polymorphs, fit residuals, and background. High-resolution TEM images of (c) $\gamma$- and (d) $\varepsilon$-WO$_3$. The inset in (c) indicates the (200) and (020) planes of $\gamma$-WO$_3$ corresponding to interlayer spacings ($d$) of 3.7 Å and 3.8 Å, respectively. The insets in (d) highlight an $\varepsilon$-WO$_3$ nanocrystal oriented along [−110] and its respective FFT. (e) DFT-relaxed structures, showing the coordination of W sites in $\gamma$-WO$_3$ (top) and $\varepsilon$-WO$_3$ with interstitial Si incorporation (bottom). Red, oxygen atom; gray, tungsten atom; blue, silicon atom. (f) W-L$_3$-edge Fourier transforms of $k^2$-weighted $\chi(k)$ together with first-shell fits, along with the (g) corresponding wavelet transform (WT, magnitude) for $\gamma$-WO$_3$. Further WT analysis is reported in Figures S4 and S5. (h) Second derivative of XANES where the edge position ($E_0$) and crystal field splitting ($\Delta$) are also indicated.

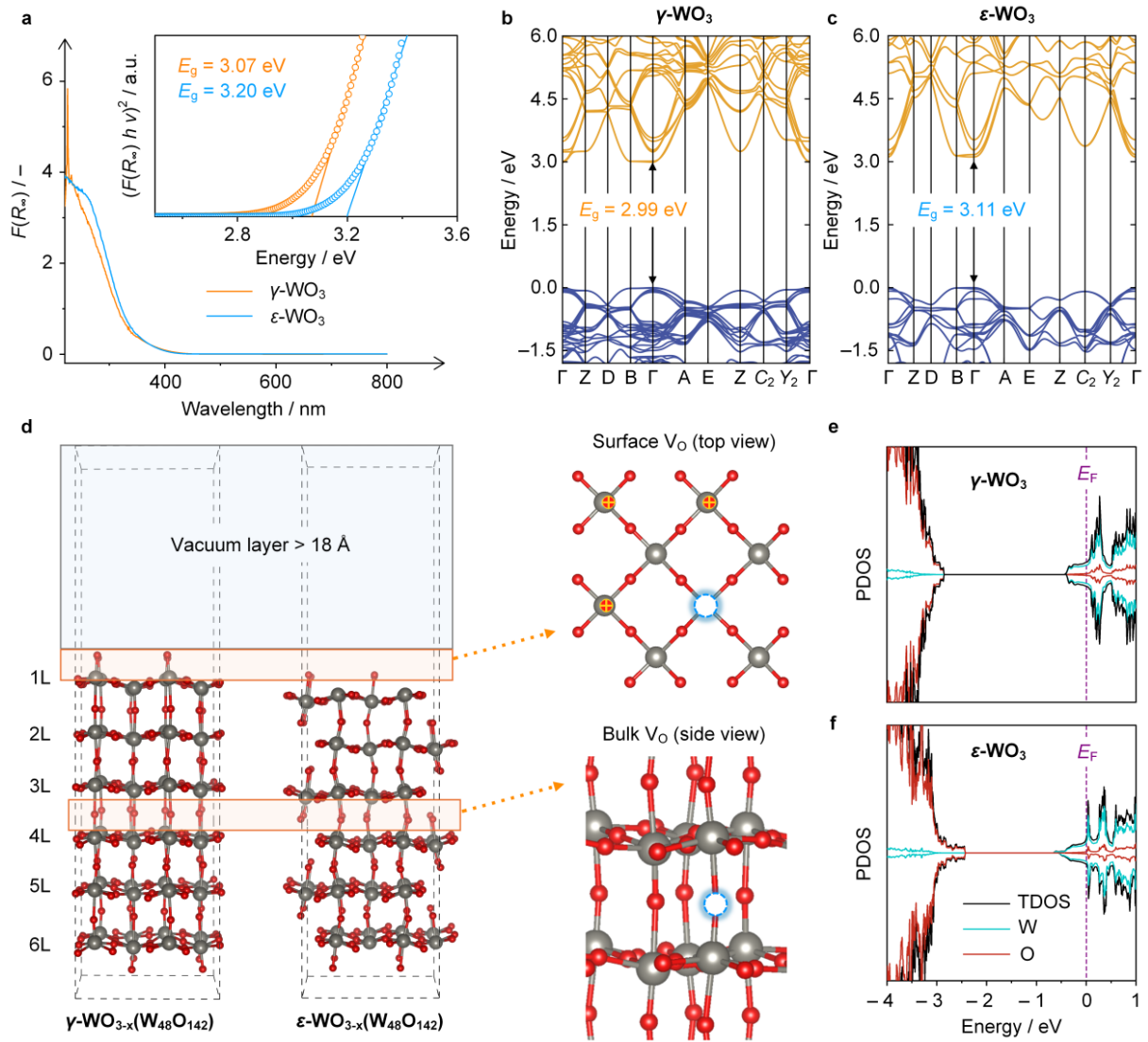

**Figure 2 | Electronic structure and atomistic model of WO$_3$.** (a) UV-Vis spectra and $E_g$ determination for $\gamma$-WO$_3$ and $\varepsilon$-WO$_3$. Calculated band structures, with valence band (blue) and conduction band (orange) states, for the unit cells of (b) $\gamma$-WO$_3$ and (c) $\varepsilon$-WO$_3$. (d) Slab models of $\gamma$-WO$_3$ (001) and $\varepsilon$-WO$_3$ (001), with a zoomed-in view illustrating the introduction of bulk and surface oxygen vacancies (V$_O$) in the $\gamma$ polymorph. Red, oxygen atom; gray, tungsten atom The highlighted O in the top view are surface atoms. (e)-(f) Projected density-of-states (PDOS) graphs for slab models of the two WO$_{3-x}$ phases with bulk and surface V$_O$.

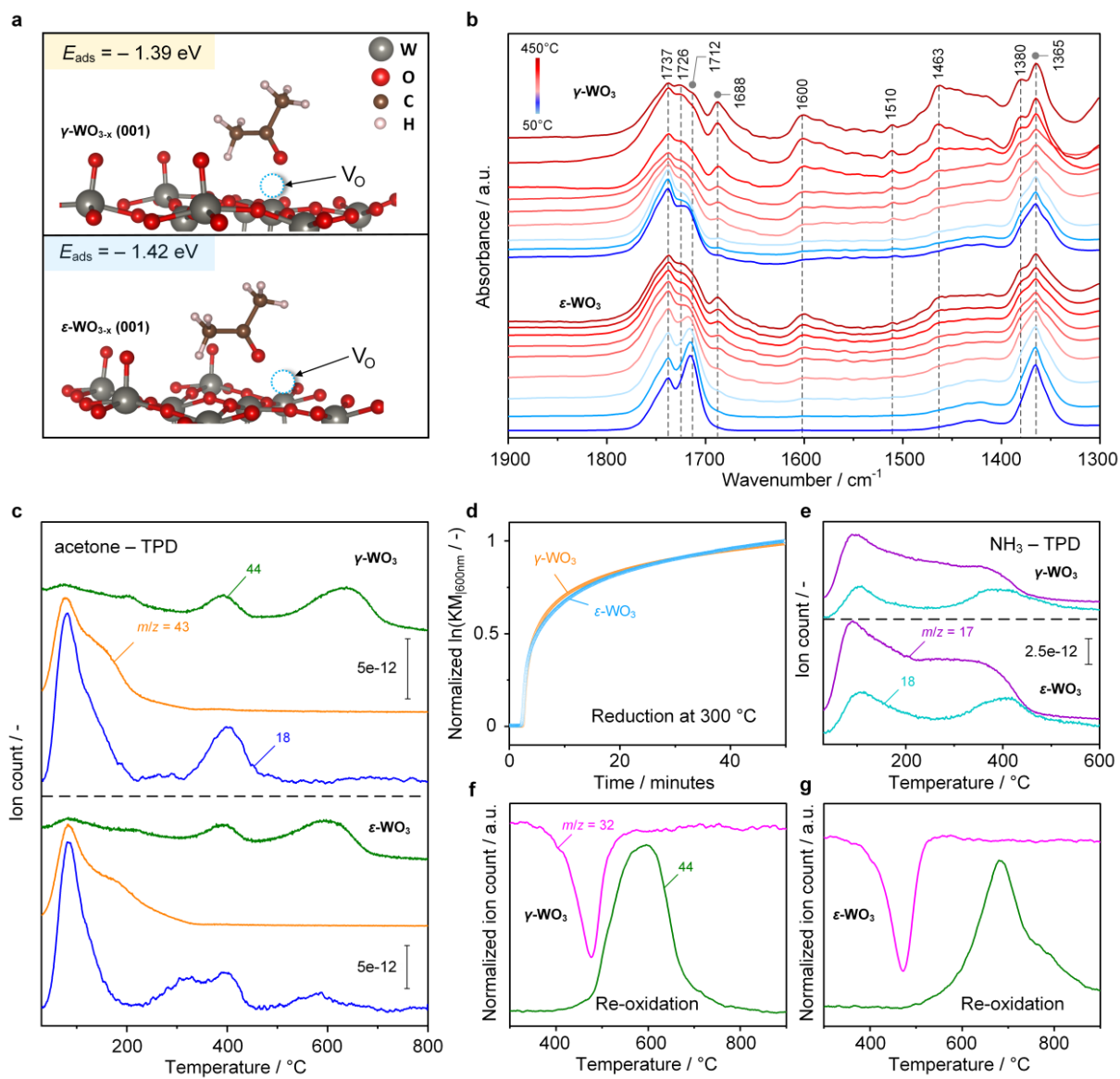

**Figure 3 | Surface chemistry with acetone and WO₃.** (a) Equilibrium structures of *γ*-WO$_{3-x}$ (001) and *ε*-WO$_{3-x}$ (001) with an adsorbed acetone molecule. Shown are the adsorption energies obtained by DFT. (b) *In situ* DRIFT spectra of adsorbed acetone between 50 – 450 °C. (c) Profiles of *m/z* = 44 (green), 43 (blue) and 18 (orange) during acetone – TPD for *γ*-WO₃ and *ε*-WO₃. (d) Transient absorbance at 600 nm (*d-d* transition of reduced W$^{(6-δ)+}$ species) measured upon exposure to acetone at 300 °C. (e) Evolution of *m/z* = 17 (purple) and 18 (cyan) curves during NH₃ – TPD. Consumption and formation of O₂ (*m/z* = 32) and CO₂ (*m/z* = 44), respectively, during O₂ – TPO following acetone – TPR (Figure S13) over (f) *γ*-WO₃ and (g) *ε*-WO₃.

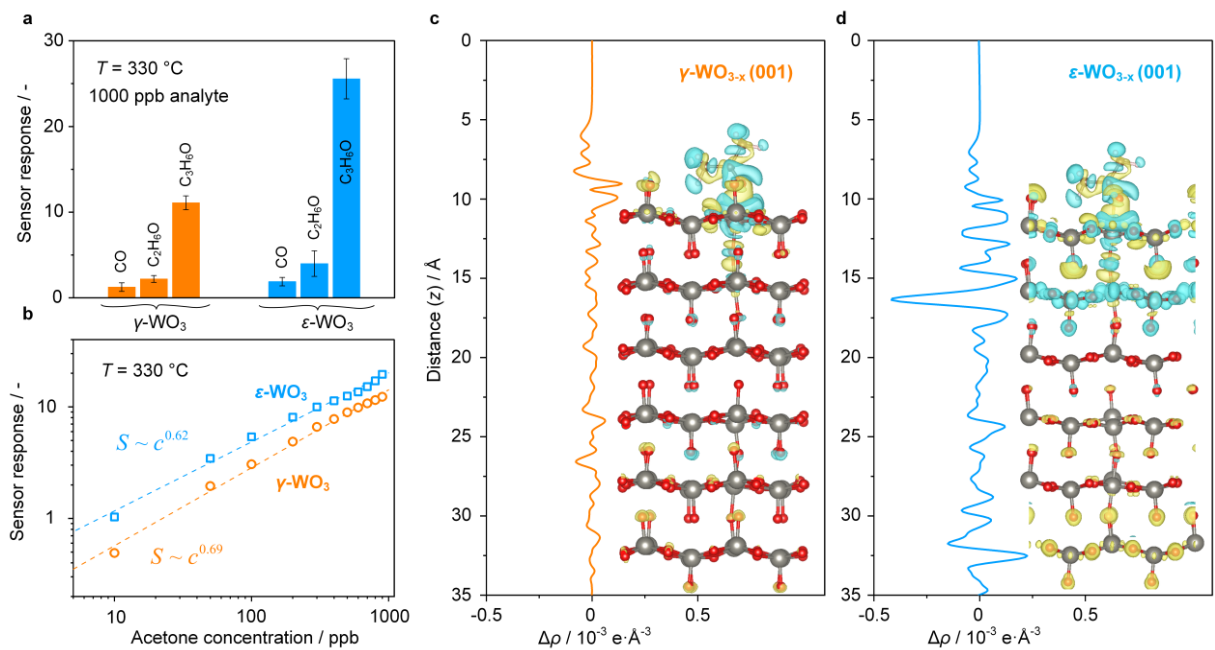

**Figure 4 | Sensing performance and lattice electronic perturbations.** (a) Sensor response of $\gamma$-WO$_3$ (orange) and $\varepsilon$-WO$_3$ (blue) towards 1000 ppb CO, ethanol and acetone, as well as (b) response vs. concentration characteristics of these for acetone ranging between 10 – 1000 ppb. The column heights and error bars are the average and standard deviation of three identically prepared sensors, operated at 330 °C and under dry air conditions. CDD maps as well as plane-averaged differential density ($\Delta\rho$) plot as a function of $z$-coordinate (i.e., normal to the slab) of acetone adsorbed on (001) surfaces of (c) $\gamma$-WO$_{3-x}$ and (d) $\varepsilon$-WO$_{3-x}$.

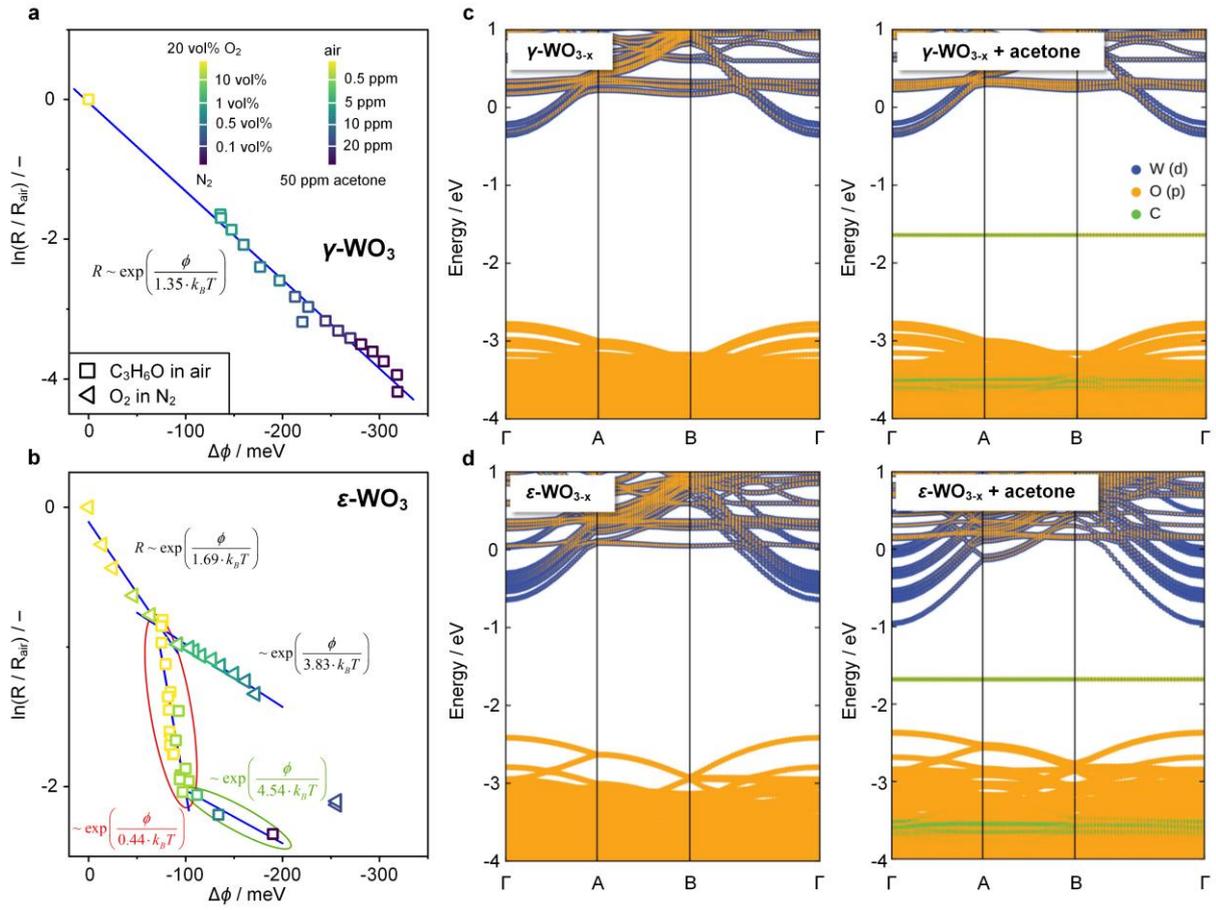

**Figure 5 | Transducer function and electronic band structure analysis.** Operando electrophysical measurements (simultaneous $\phi$ and $R$) over (a) $\gamma$- and (b) $\varepsilon$-WO$_3$, under 0 – 20 vol% O$_2$/N$_2$ (triangles) and 0 – 50 ppm C$_3$H$_6$O/air (squares). Electronic band structure analysis of (c) $\gamma$- and (d) $\varepsilon$-WO$_{3-x}$ (001) before and after adsorption of an acetone molecule.

# Supplementary Information

**Supplementary Note S1**

To complement the DRIFTS analysis in Figure 3b, we conducted AIMD simulations of acetone adsorbed near a surface $V_O$ on $\gamma$-WO$_{3-x}$ (001) at 500 K (Figure S9). The trajectory confirms that acetone remains stably bound over 10 ps, supporting the thermal robustness of the adsorption complex under sensing-relevant conditions. The C=O bond length ($d_{C-O}$) fluctuates dynamically as shown in Figure S10, with transient elongations beyond 1.30 Å (Figure S10), indicative of vibrational softening and C=O activation ($d_{C-O}^{initial} = 1.246 \text{ Å}$). Beyond C=O dynamics, AIMD also reveals pronounced torsional motion of the methyl groups, particularly in the form of hindered rotations and out-of-plane bending, likely contributing to the broader spectral $\delta$(CH$_3$)-shoulder near 1380 cm$^{-1}$.

**Supplementary Note S2**

To assess lattice reducibility under acetone exposure, we monitored the evolution of KM function at 600 nm, attributed to the *d-d* transitions[81] of reduced W$^{(6-\delta)+}$ species. Both polymorphs exhibit progressive increase in signal intensity and similar $k_{red}$ of ~ 0.14 min$^{-1}$ (from the initial KM slope[79]), reflecting electron transfer to coordinatively unsaturated tungsten, in agreement with the Bader charge redistribution. NH$_3$-TPD profiles (Figure 3e) indicate the presence of such Lewis-acidic W sites via *m/z* = 17 desorption that are uncorrelated with and exceeding signal intensity of *m/z* = 18,[82] excluding simple H$_2$O evolution. Py chemisorption (Figure S11) further confirms the availability of these Lewis sites, with both polymorphs retaining Py up to 350 °C. Thereby, $\varepsilon$-WO$_3$ exhibits a slightly higher desorption onset, indicating comparable but marginally stronger acid-base interactions.

Following acetone-reduction, O$_2$-TPO (Figure 3f,g) reveals a significant O$_2$-uptake between 400 – 550 °C for both polymorphs. Subsequent CO$_2$ evolution at higher temperatures reflects the oxidative removal of residual carbon species, indicating that acetone-derived intermediates remained on the surface and required replenished $V_O$ to be fully reacted. Under CO- and $H_2$-TPR conditions (Figure S12), $V_O$ formation after lattice oxygen consumption occurs at rather high temperatures (> 500 °C). In contrast, $C_3H_6O$-TPR (Figure S13) features measurable redox activity near 400 °C, accompanied by the formation of surface-bound intermediates (e.g., enolates, formates, condensation products) that may undergo deep dehydrogenation at higher temperatures, as evidenced by the pronounced $H_2$ release ($m/z = 2$) above 600 °C. This lower reducibility aligns with the characteristically weak CO and $H_2$ responses observed[45] in $WO_3$-based chemoresistors. Finally, $m/z = 44$ evolution during $CO_2$-TPD (Figure S14) shows lower intensity and a distinct thermal onset compared to Figure 3c, indicating[74] that acetone more readily mobilizes lattice oxygen — in agreement with the MvK-type behavior evidenced across Figure 3c–g.

**Supplementary Note S3**

The absence of monotonic increases in both $\varphi$ and $R$ with rising $O_2$ concentration (Figure S19) is attributed to the dominant role of molecularly adsorbed $O_2$, i.e., $O_2$(ad), on the surface of $\gamma$-$WO_3$. By contrast, $\varepsilon$-$WO_3$ displays a concurrent rise in both electrophysical parameters (Figure S22), yielding a piecewise linear transduction behavior (on a semi-log scale, Figure 5b) consistent with *operando* $\varphi$ measurements. In fact, such $O_2$(ad) species are more abundant on $\gamma$- rather than on $\varepsilon$-$WO_3$ as supported by stronger binding (lower $E_{ads}(O_2)$, Figure S25) and more pronounced low-temperature $O_2$ desorption[88] (Figure S26). Unlike strongly (iono-)sorbed[89] $O_\beta^{\alpha-}$ that are also present on $\gamma$-$WO_3$ (monotonic $R$-increase in Figure S19), $O_2$(ad) do not withdraw CB electrons and therefore do not contribute to build up a delocalized space-charge width.[90] Instead, they can form localized dipoles that alter $\varphi$ via electron affinity ($\chi$) shifts,[90-92] partially counteracting—i.e., $\delta^-$ charge closer to the surface[93]—the upwards $O_\beta^{\alpha-}$-related band bending ($qV_s$).

**Supplementary Note S4**

The SnO$_2$ precursor was prepared by dissolving tin(II) 2-ethylhexanoate (90 wt% in 2-ethylhexanoic acid, Sigma Aldrich) in xylene (ACS Reagent, Sigma Aldrich) to obtain a Sn molarity of 0.2 mol L$^{-1}$. The precursor was fed to a spray flame with the same process parameters as for $\gamma$- and $\varepsilon$-WO$_3$, and SnO$_2$ sensors were similarly obtained by direct deposition (4 minutes) from the aerosol onto identical sensor substrates positioned at 20 cm above the nozzle. After annealing in air (5 hours at 500 °C), they were mounted onto MACOR® holders and measured in our *operando* work function ($\varphi$) setup (see Methods).

Figures S26 and S27 show simultaneous $\varphi$ and resistance traces of SnO$_2$ operated at 400 °C during exposure to 8 – 175 ppm CO and 0.5 – 16.5 ppm H$_2$ in dry air. In both cases, the measured resistance falls between its values in synthetic air and N$_2$, consistent with operation under an electron-depletion regime,[94] as expected for SnO$_2$ in high-oxygen (~ 20 vol%) backgrounds. As shown in Figure S30, the $R$ vs. $\varphi$ data for H$_2$/air (green) and CO/air (red) collapse onto a single regression line with an inverse coefficient of 1.19, in excellent agreement with the literature.[95] Notably, this value is close to that obtained under O$_2$/N$_2$ at 300 °C (inverse coefficient of 1.27, Figure S29), validating our *operando* methodology. At 400 °C, however, O$_2$/N$_2$ exposures may have led to electron affinity ($\chi$) shifts (thus, they are not presented), similar to those discussed for $\gamma$-WO$_3$ (see Supplementary Note S3 and Figure S19), reflecting temperature-dependent surface dipole contributions.[90]

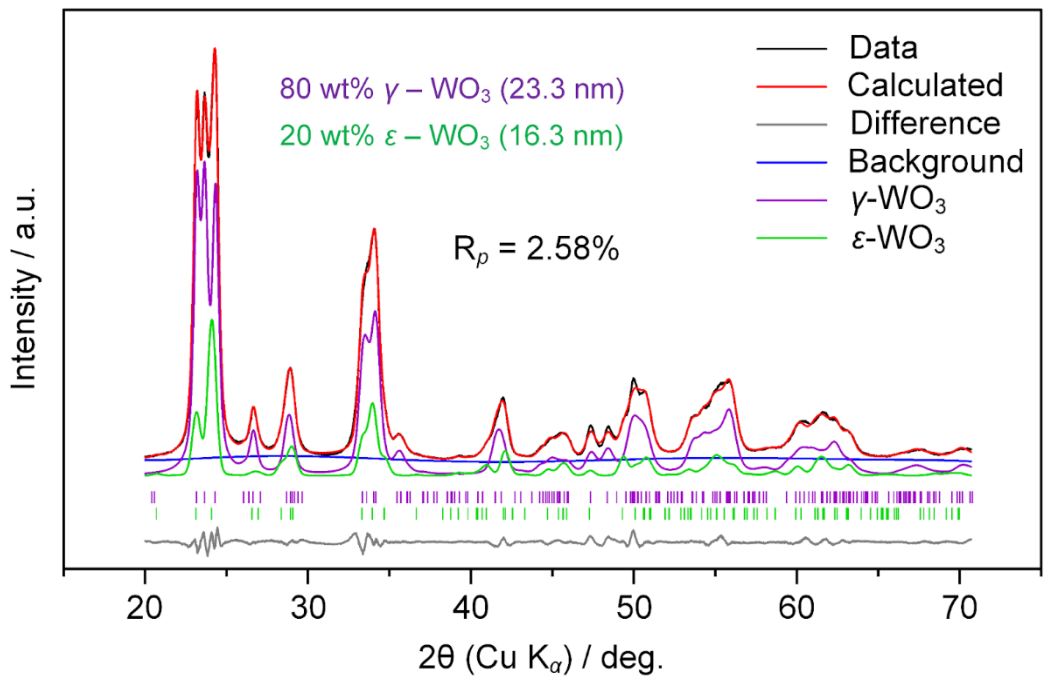

**Figure S1:** XRD refinement of pure $WO_3$ nanocrystals accounting for the presence of $\varepsilon$-$WO_3$.

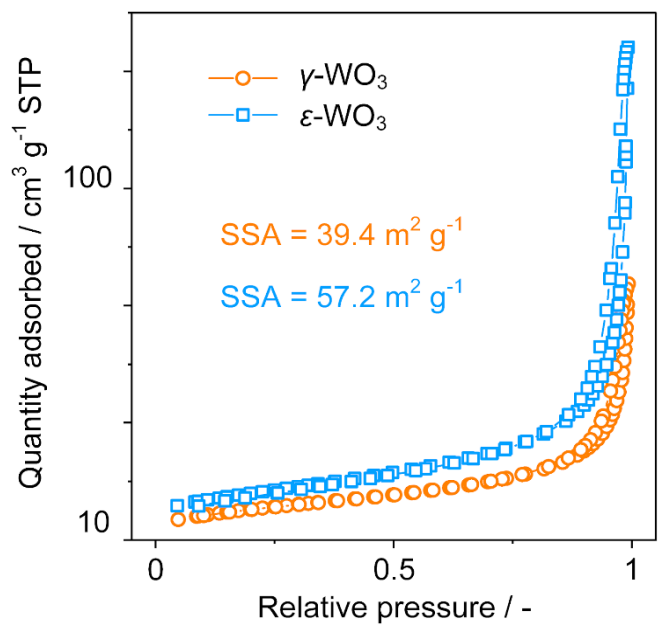

**Figure S2:** $N_2$-adsorption isotherm of $\gamma$-$WO_3$ and $\varepsilon$-$WO_3$.

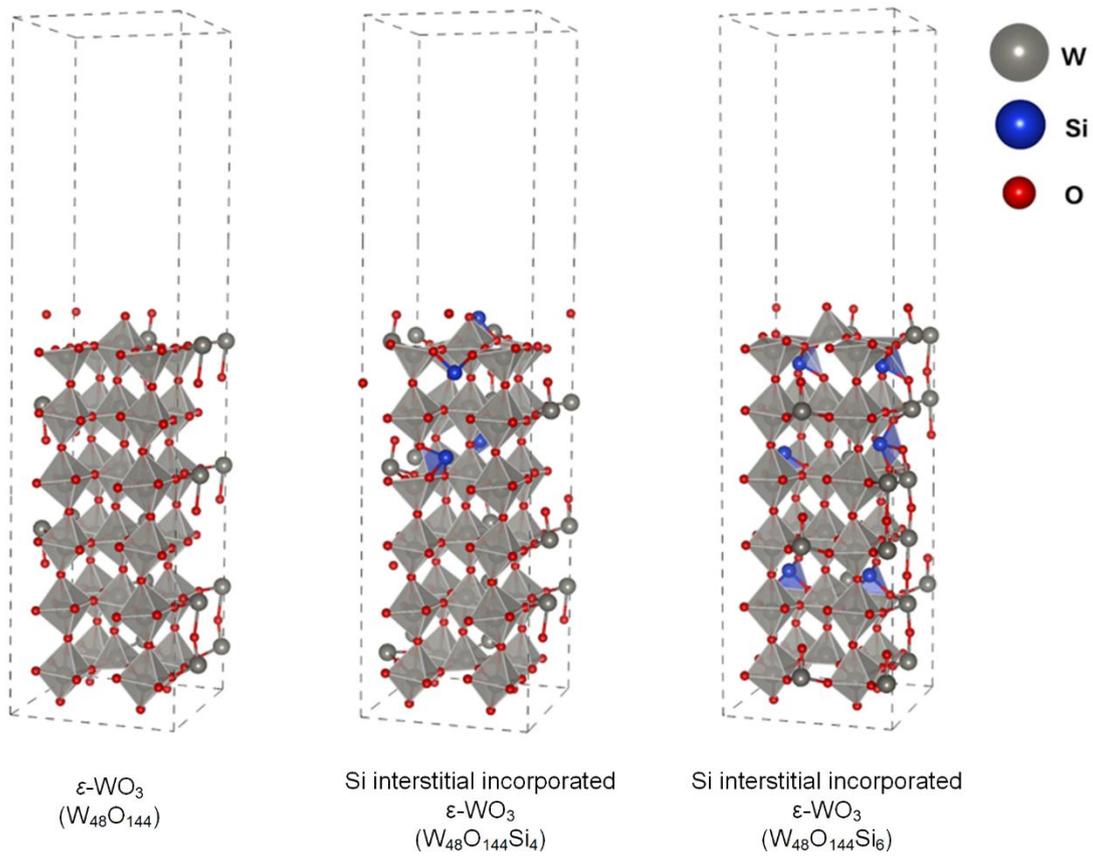

ε-WO₃
(W₄₈O₁₄₄)

Si interstitial incorporated
ε-WO₃
(W₄₈O₁₄₄Si₄)

Si interstitial incorporated
ε-WO₃
(W₄₈O₁₄₄Si₆)

**Figure S3:** DFT calculated pure and interstitial Si-incorporated $\varepsilon$-WO$_3$ (001) at varying concentrations.

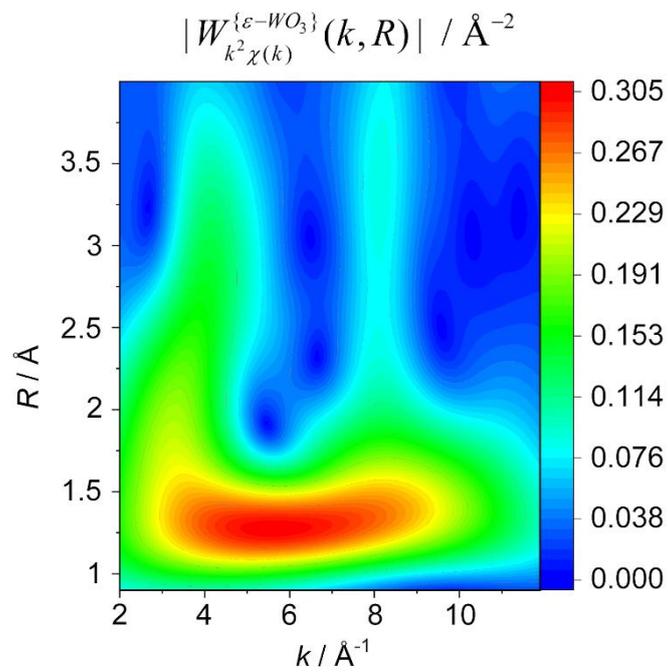

**Figure S4:** Magnitude of wavelet transform of $k^2\chi(k)$ at the W-L$_3$ edge for $\varepsilon$-WO$_3$.

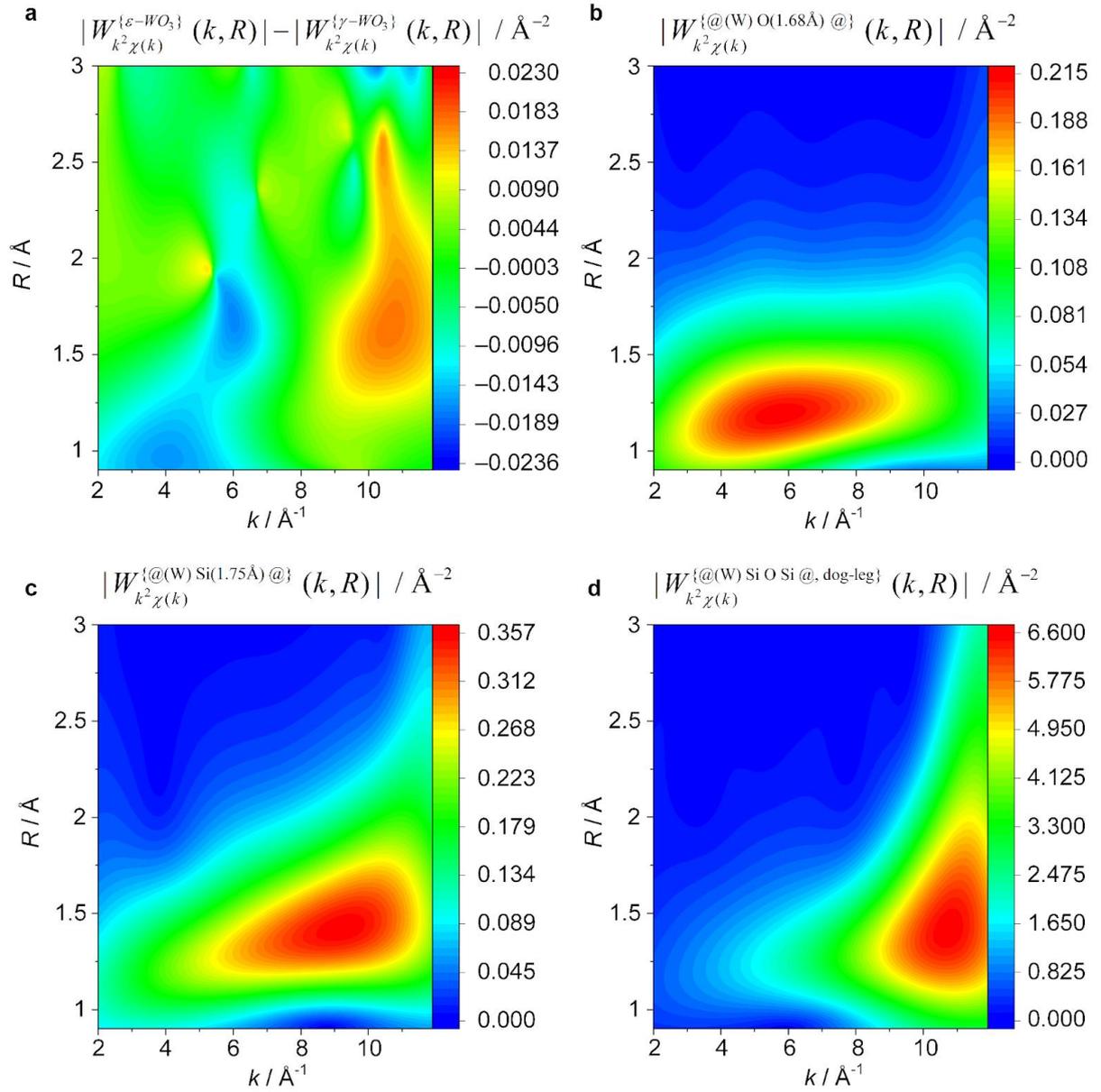

**Figure S5:** (a) Wavelet-transform difference map between $\varepsilon$-WO$_3$ and $\gamma$-WO$_3$ samples. Theoretical magnitudes of the wavelet transform of $k^2\chi(k)$ from specific scattering paths: (b) single W – O scattering, (c) single W – Si scattering as well as (d) multiple W – Si – O – Si, "dogleg"-like, scattering. All wavelet transforms were computed for $k$ and $R$ between 2 – 12 Å$^{-1}$ and 0.9 – 4 Å, respectively, with the same Morlet's wavelet parameters ($\eta$ = 5 and $\sigma$ = 1).

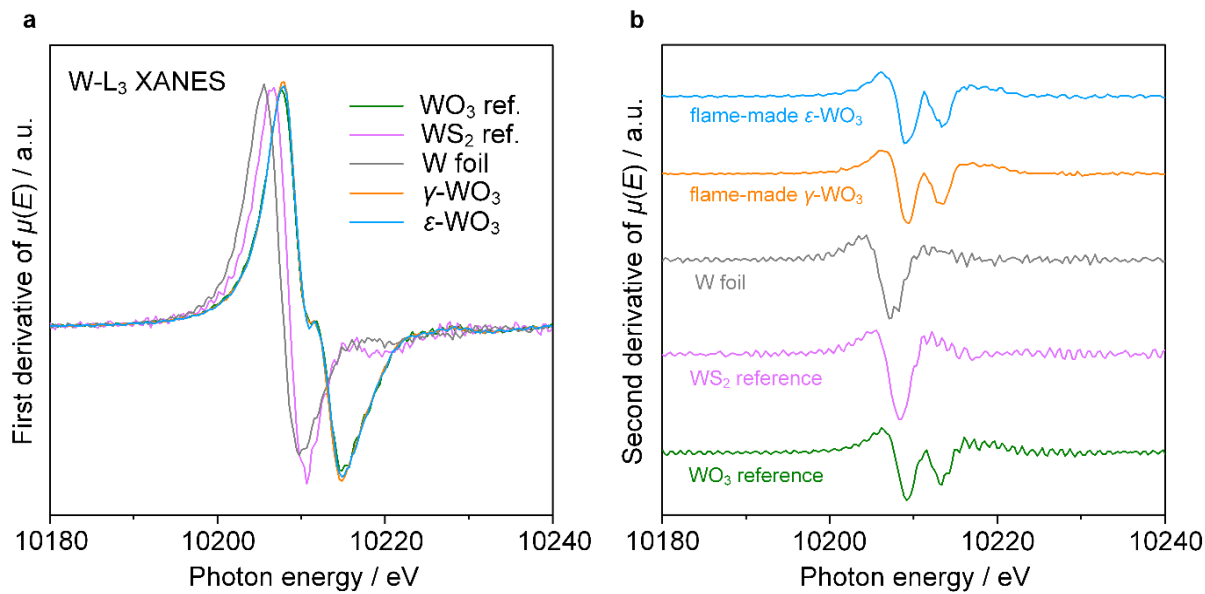

**Figure S6:** (a) First derivative of the W-L$_3$ XANES for $\gamma$-WO$_3$, $\varepsilon$-WO$_3$, as well as for the reference WO$_3$, WS$_2$, and the W foil.

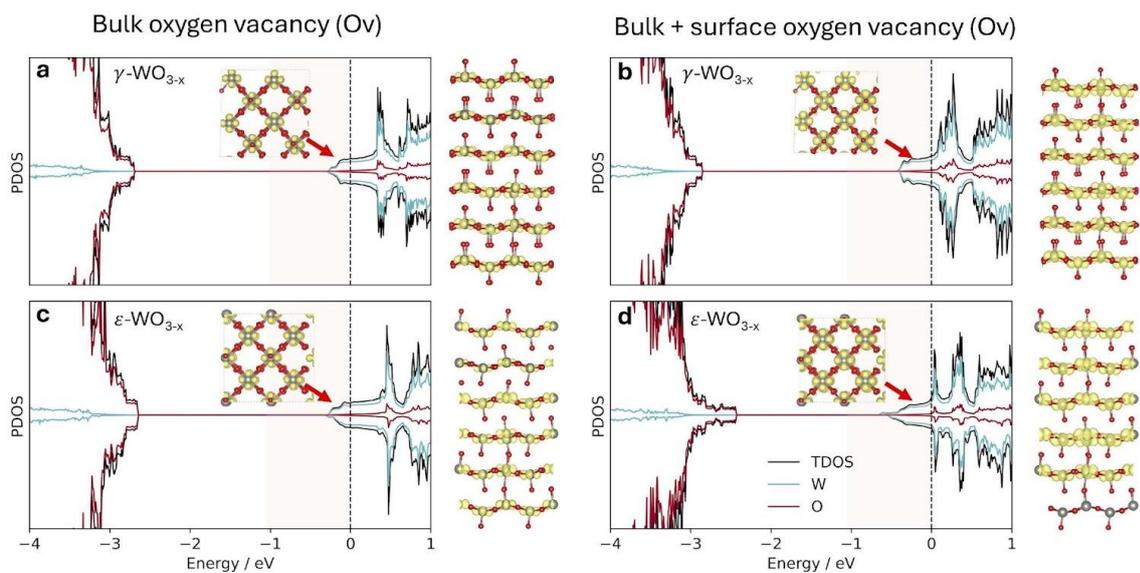

**Figure S7:** Projected density of states (PDOS) and partial charge density: the inset and right plots in each figure illustrate the electron density within 1 eV below $E_\text{F}$, with an isosurface level of $0.001\ e \cdot$ Å$^{-3}$.

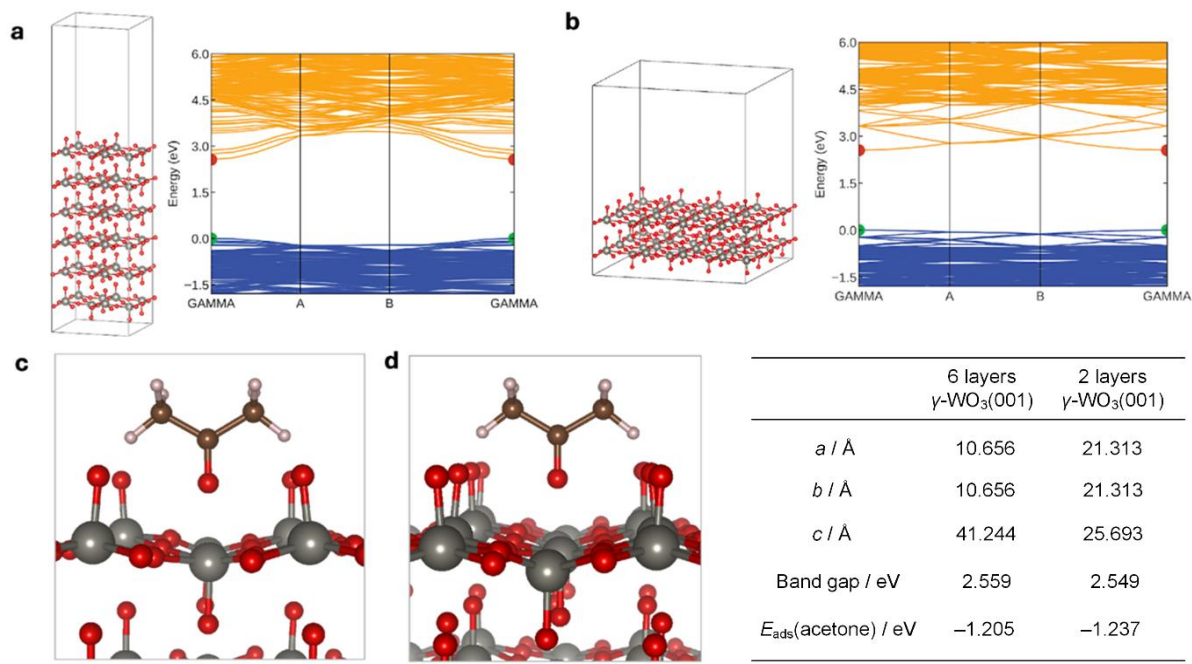

**Figure S8:** Size effects on WO$_3$'s electronic properties and acetone adsorption: (a) 6-layer γ-WO$_3$ (001) and its band structure. (b) 2-layer γ-WO$_3$ (001) and its band structure. (c-d) Optimal acetone adsorption structures on 6- and 2-layer γ-WO$_3$ (001). The accompanying table summarizes calculation details, including lattice constants, band gaps, and acetone adsorption energies ($E_{ads}$).

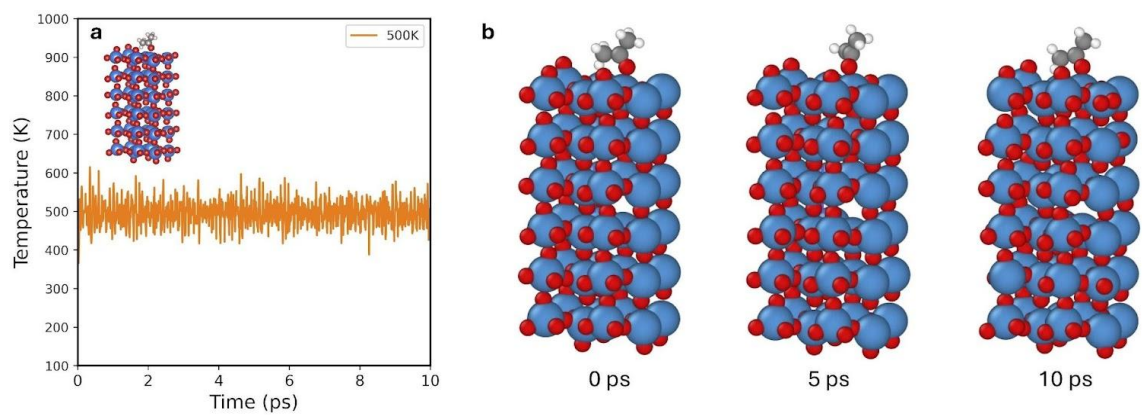

**Figure S9:** AIMD simulation of γ-WO$_3$ (001) + acetone at 500 K: (a) Temperature stabilization over 10 ps. (b) Snapshots of the acetone-adsorbed structure at 0, 5, and 10 ps.

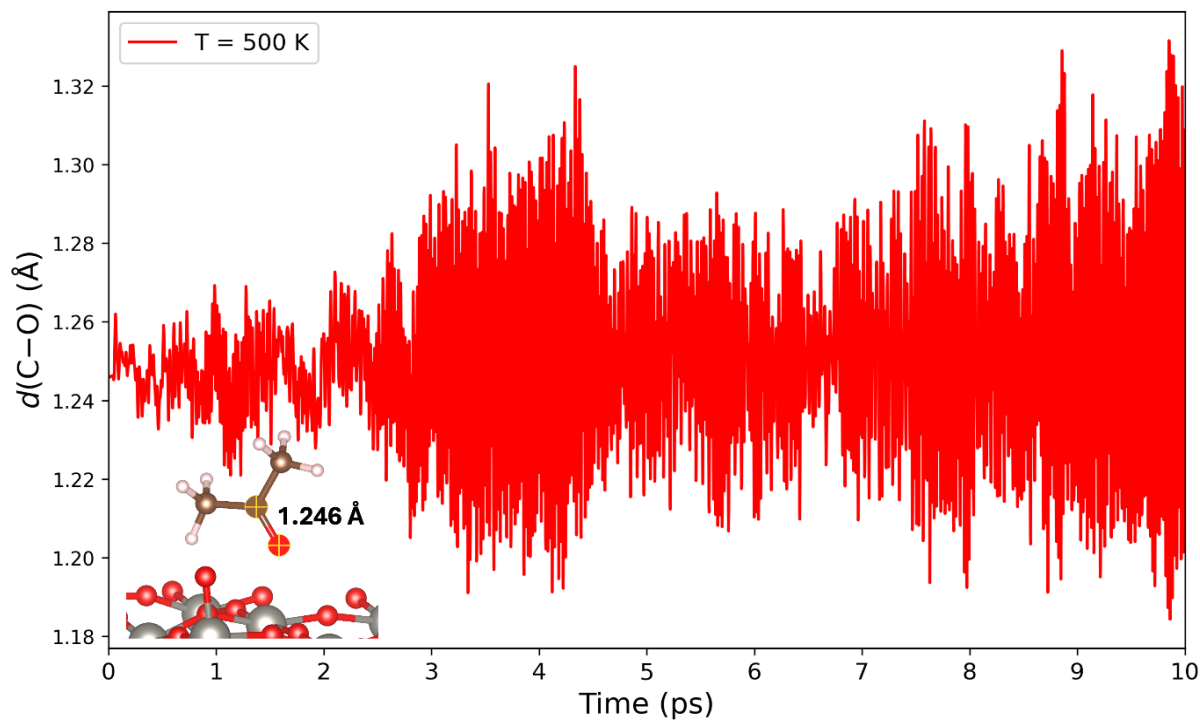

**Figure S10:** Time-trace of C=O bond length during AIMD of acetone adsorbed on $\gamma$-WO$_{3-x}$ (001).

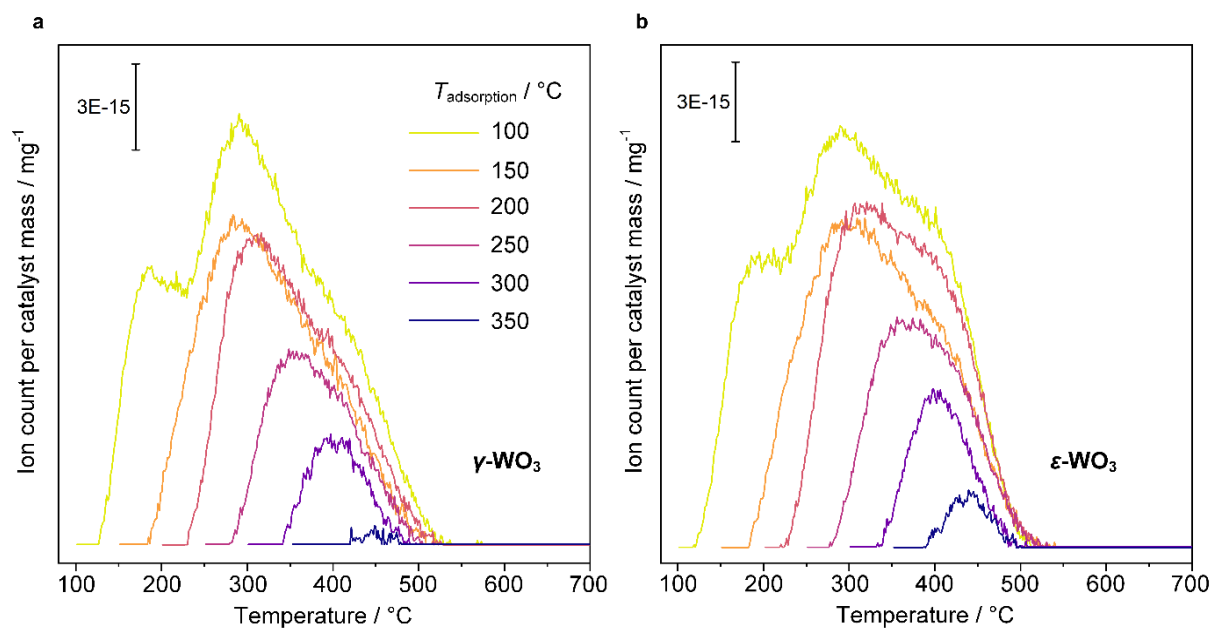

**Figure S11:** Pyridine (Py) evolution from (a) $\gamma$-WO$_3$ and (b) $\varepsilon$-WO$_3$ monitored at $m/z = 79$ during Py-TPD, following Py-adsorption at different temperatures between 100 – 350 °C.

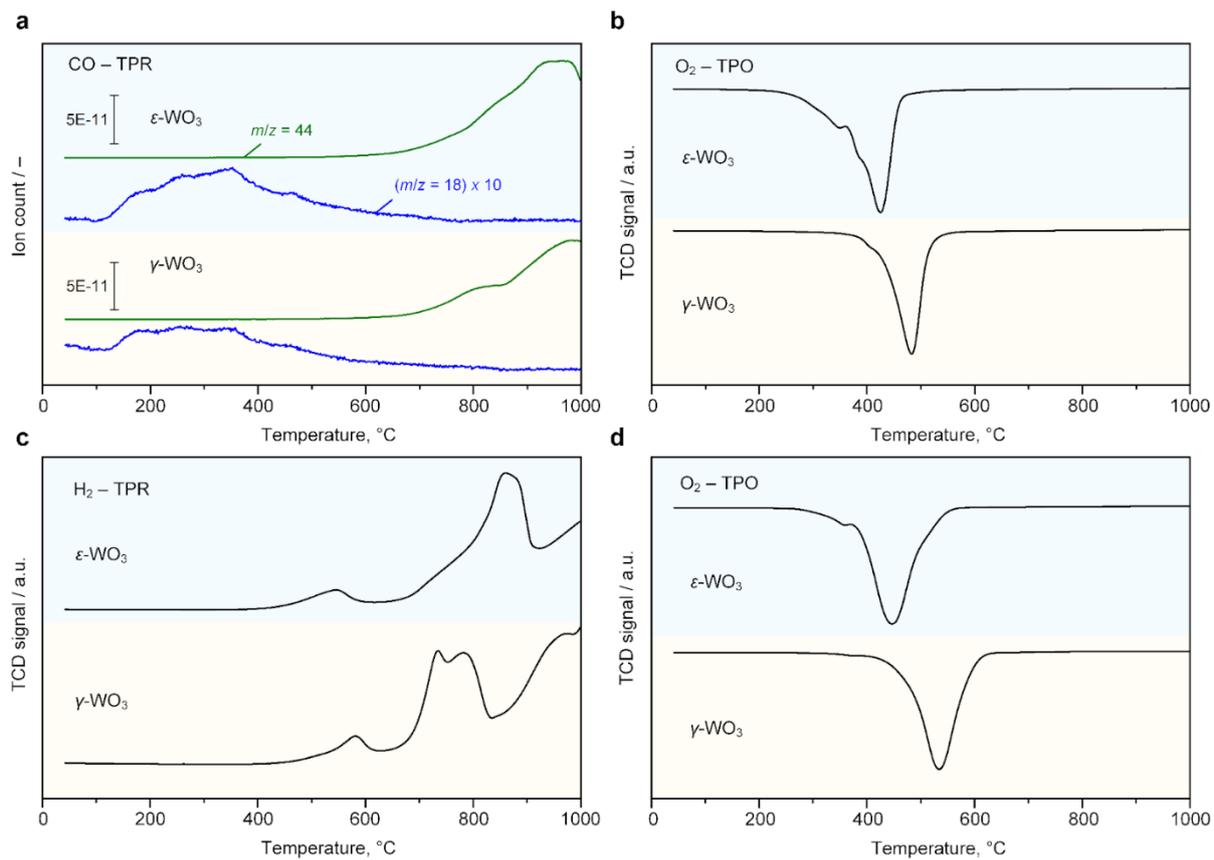

**Figure S12:** (a) CO-TPR followed by (b) $O_2$-TPO as well as (c) $H_2$-TPR followed by (d) $O_2$-TPO. Note that, in (a), the *m/z* = 18 trace is scaled (*x*10) for both polymorphs.

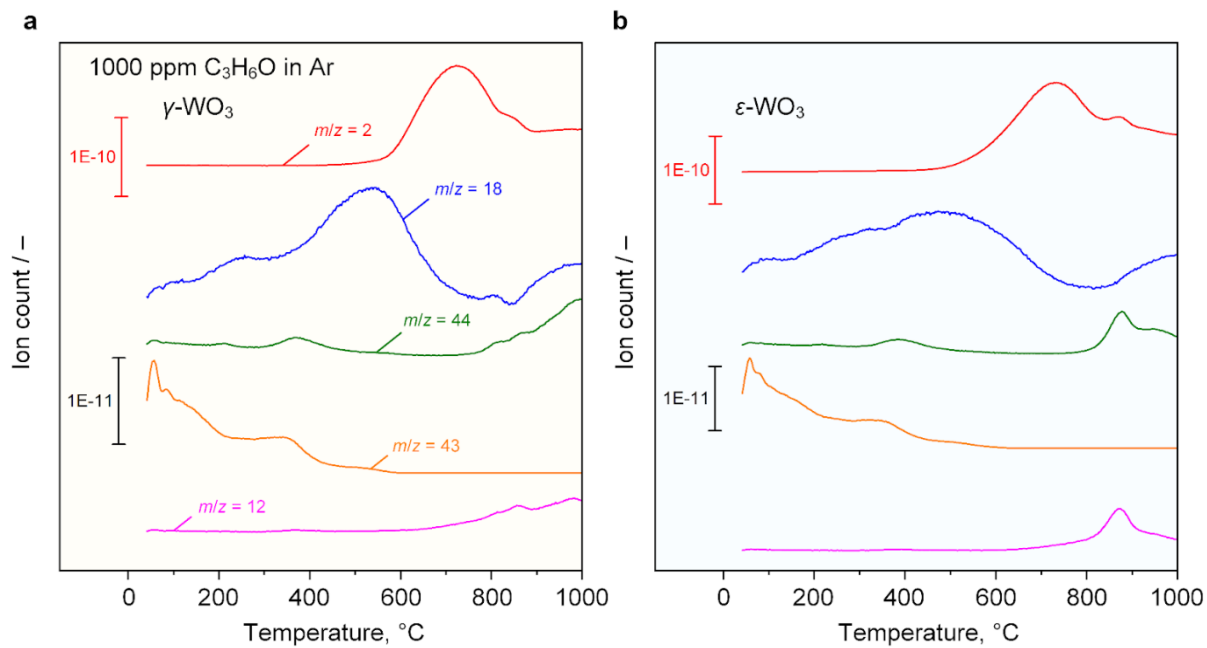

**Figure S13:** Evolution of $m/z$ = 2, 18 and 44 during $C_3H_6O$-TPR performed over (a) $\gamma$-$WO_3$ and (b) $\varepsilon$-$WO_3$. Note that, in both (a) and (b), the $m/z$ = 2 signal and $m/z$ =12, 18, 43 and 44 are referenced to different scale bars of 1e-10 and 1e-11, respectively.

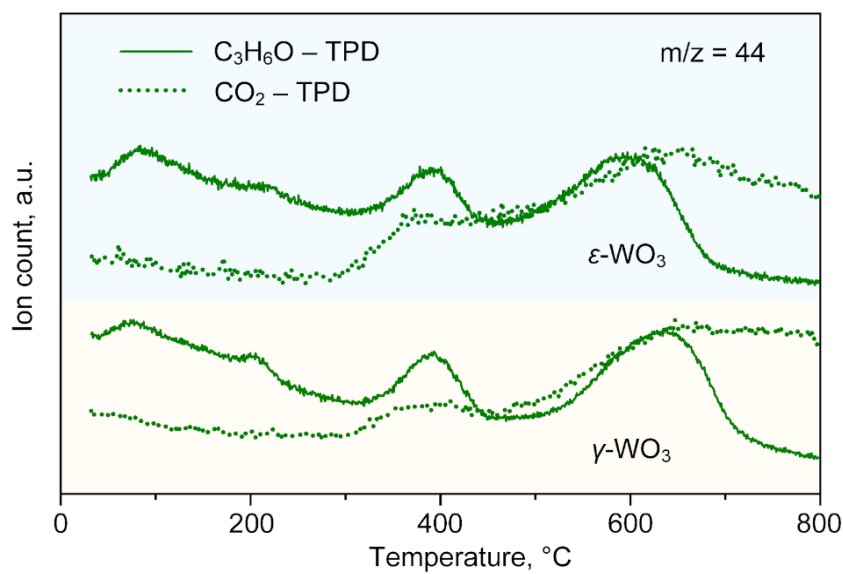

**Figure S14:** $CO_2$-evolution ($m/z$ = 44) during $C_3H_6O$-TPD and $CO_2$-TPD over $\gamma$-$WO_3$ and $\varepsilon$-$WO_3$.

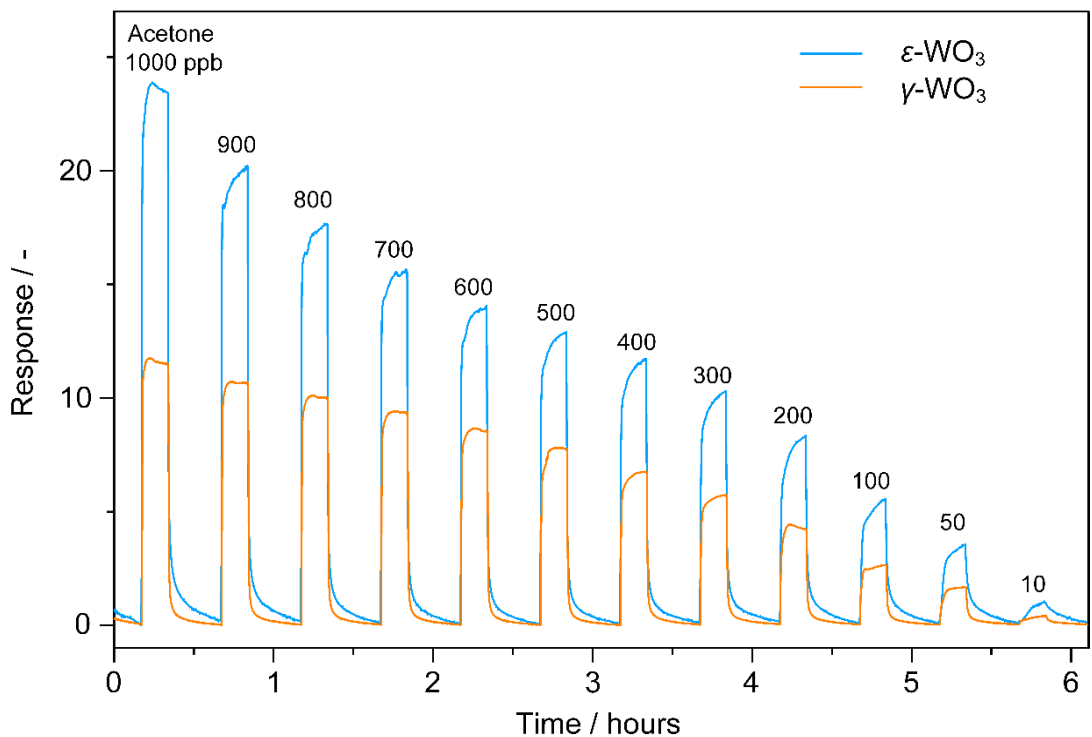

**Figure S15:** Sensor response transients of $\gamma$-WO$_3$ and $\varepsilon$-WO$_3$ upon exposure to 10 – 1000 ppb acetone at 330 °C in dry air.

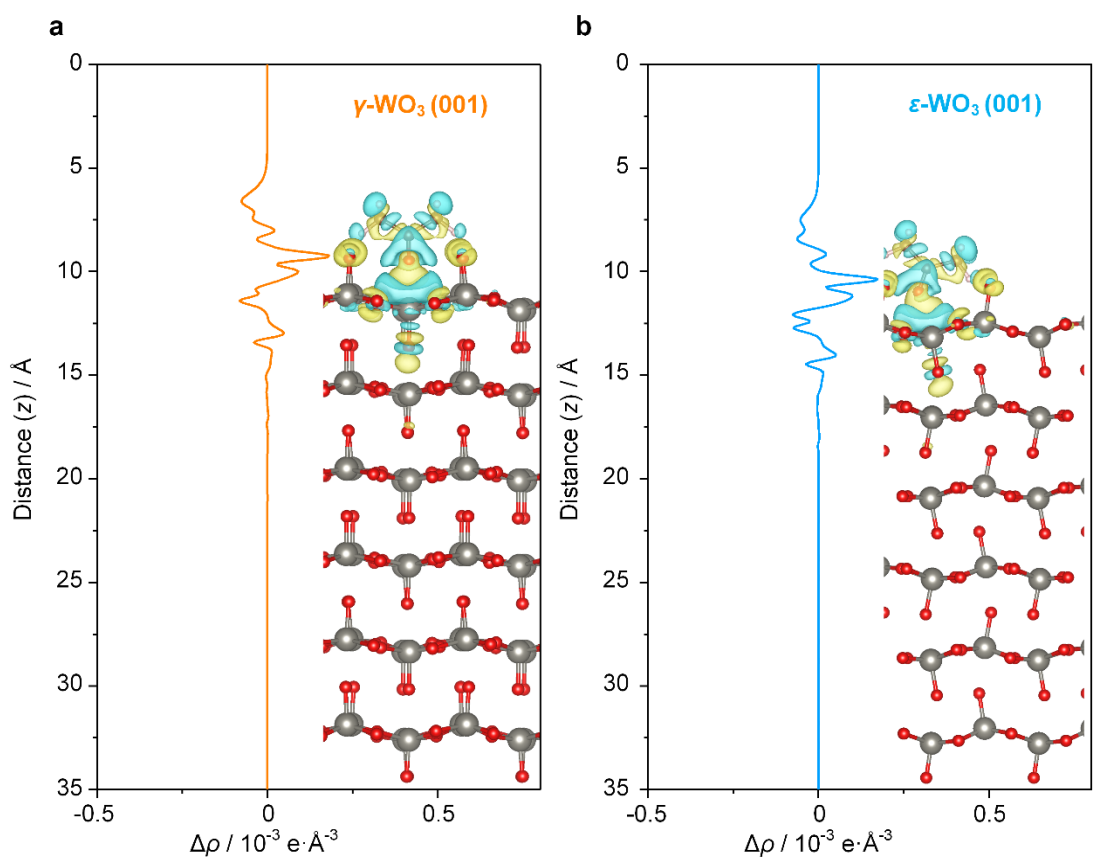

**Figure S16:** $\Delta\rho(z)$ along with CDD map of acetone adsorbed onto stoichiometric (a) $\gamma$- and (b) $\varepsilon$-WO$_3$.

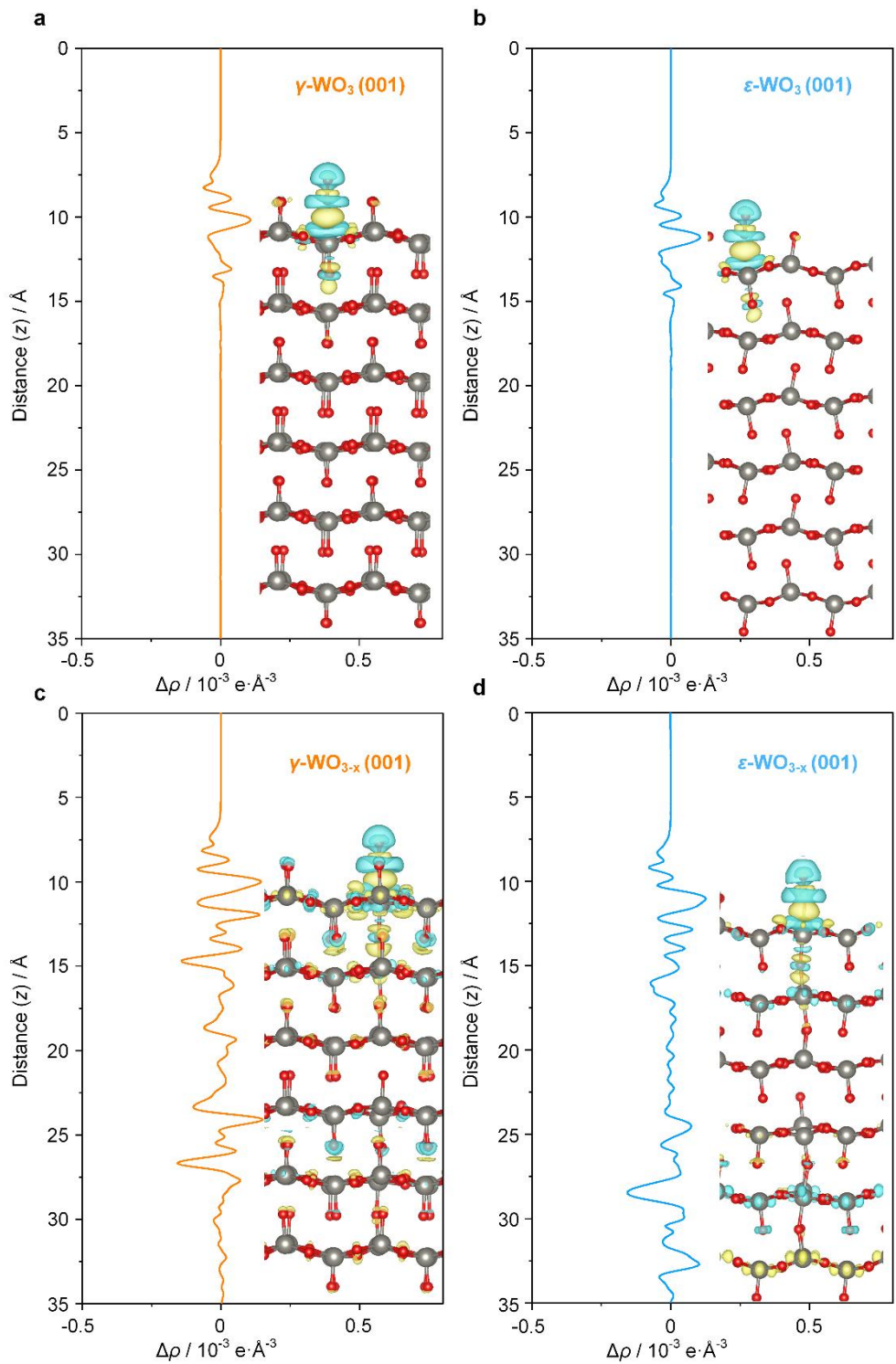

**Figure S17:** CDD maps of CO adsorbed on stoichiometric (a) $\gamma$– and (b) $\varepsilon$–WO$_3$, as well as reduced (c) $\gamma$– and (d) $\varepsilon$–WO$_{3-x}$, along with their respective plane-averaged $\Delta\rho(z)$.

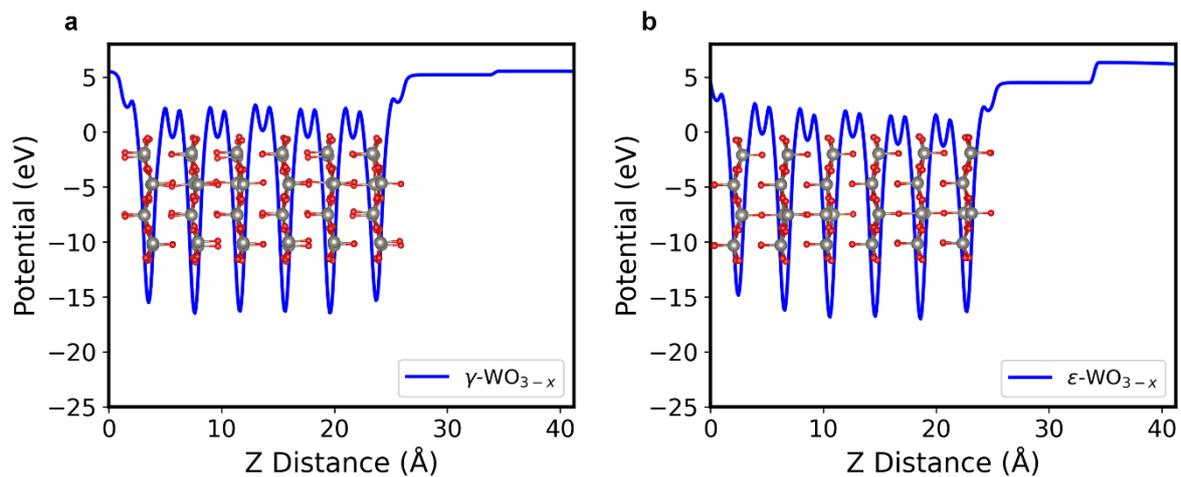

**Figure S18:** Electrostatic potential profiles across (a) $\gamma$- and (b) $\varepsilon$-WO$_{3-x}$ (001) slabs.

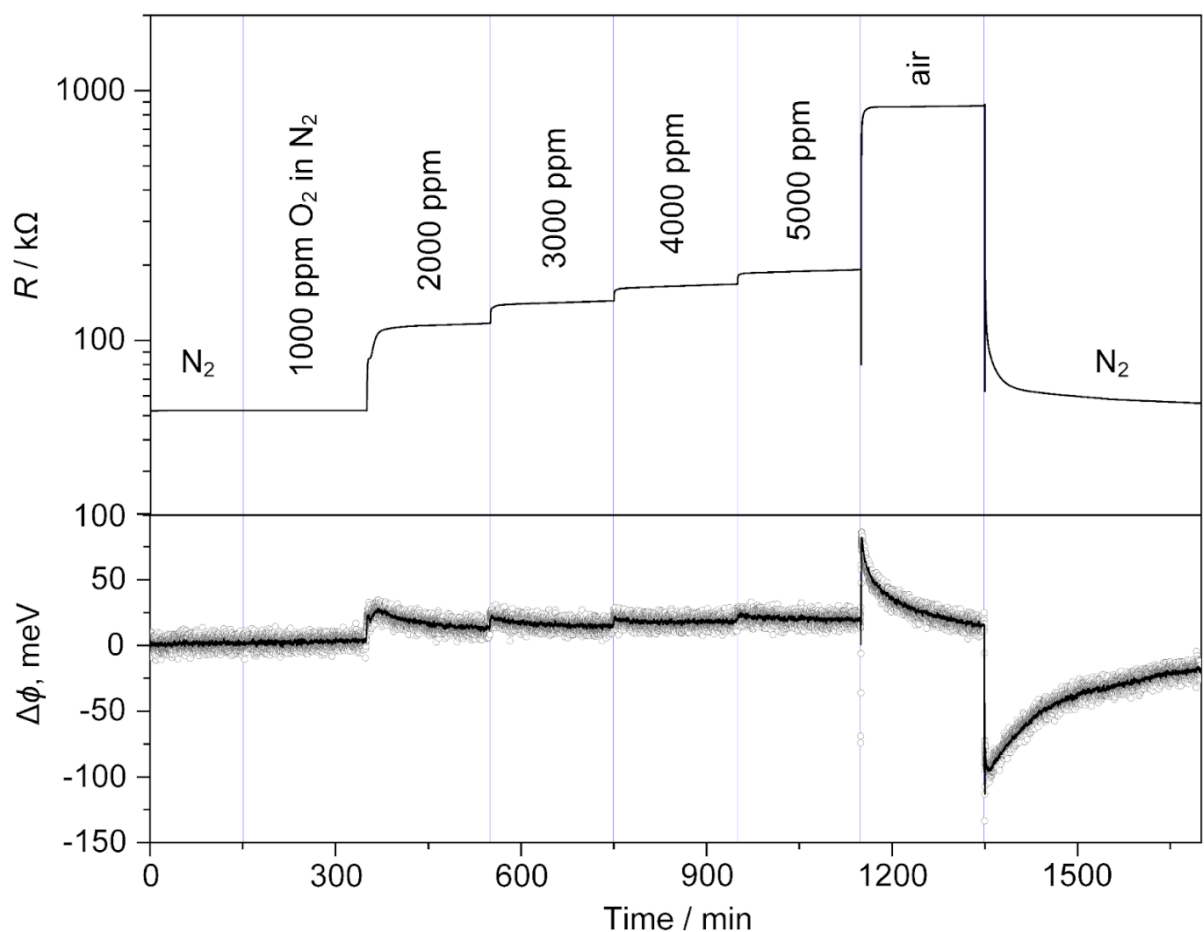

**Figure S19:** *Operando* work function measurements upon exposing $\gamma$-WO$_3$ to 0 – 20 vol% O$_2$ in N$_2$ at 330 °C.

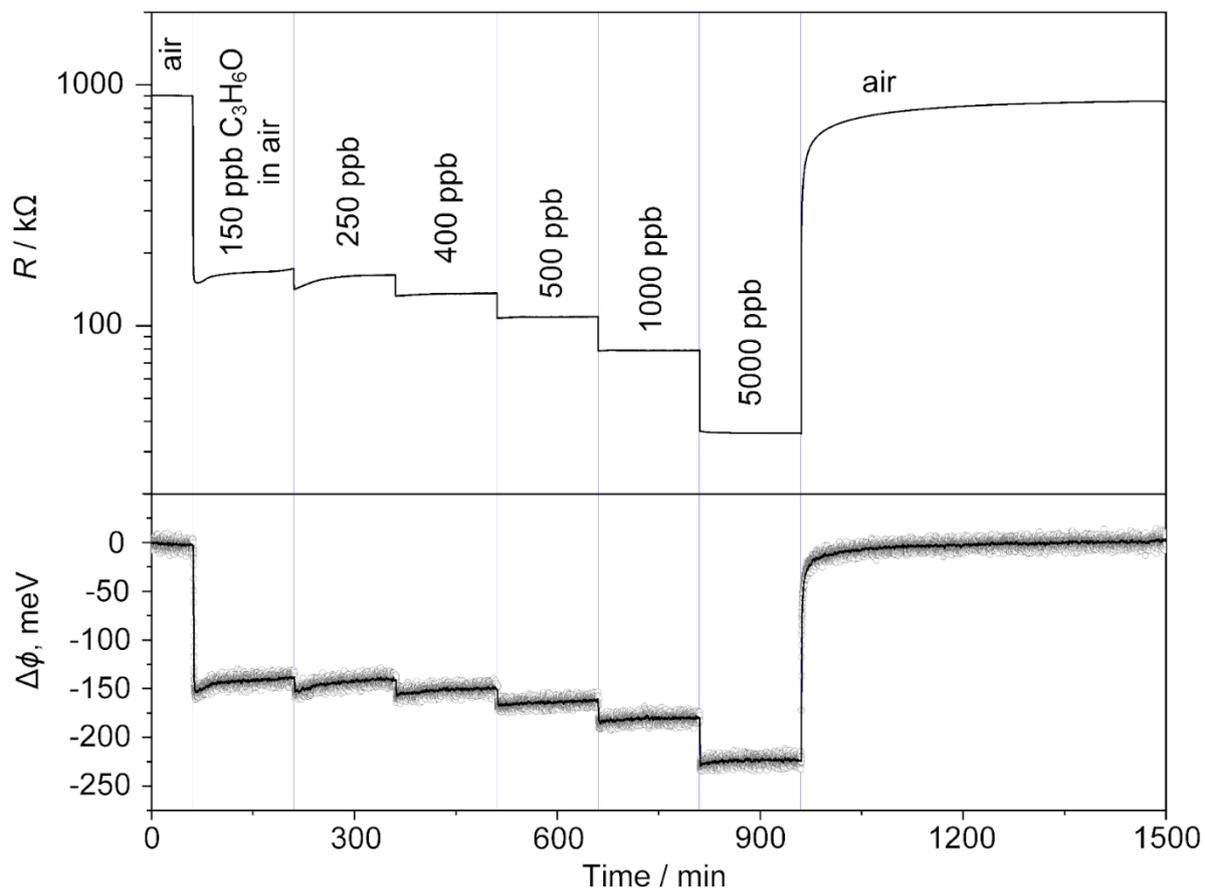

**Figure S20:** *Operando* work function measurements upon exposing $\gamma$-$WO_3$ to 150 – 5000 ppb $C_3H_6O$ in air at 330 °C.

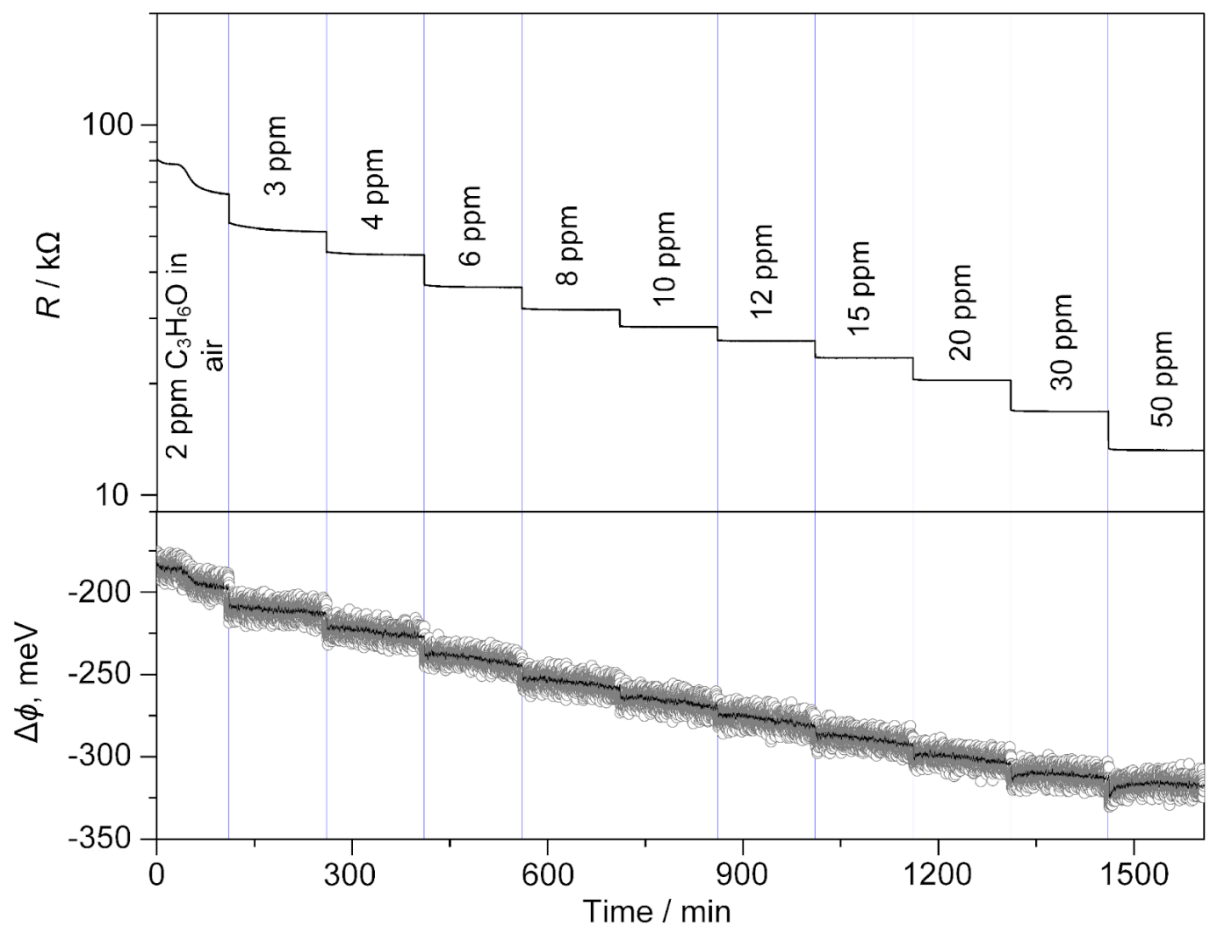

**Figure S21:** *Operando* work function measurements upon exposing $\gamma$-WO$_3$ to 2 – 50 ppm C$_3$H$_6$O in air at 330 °C.

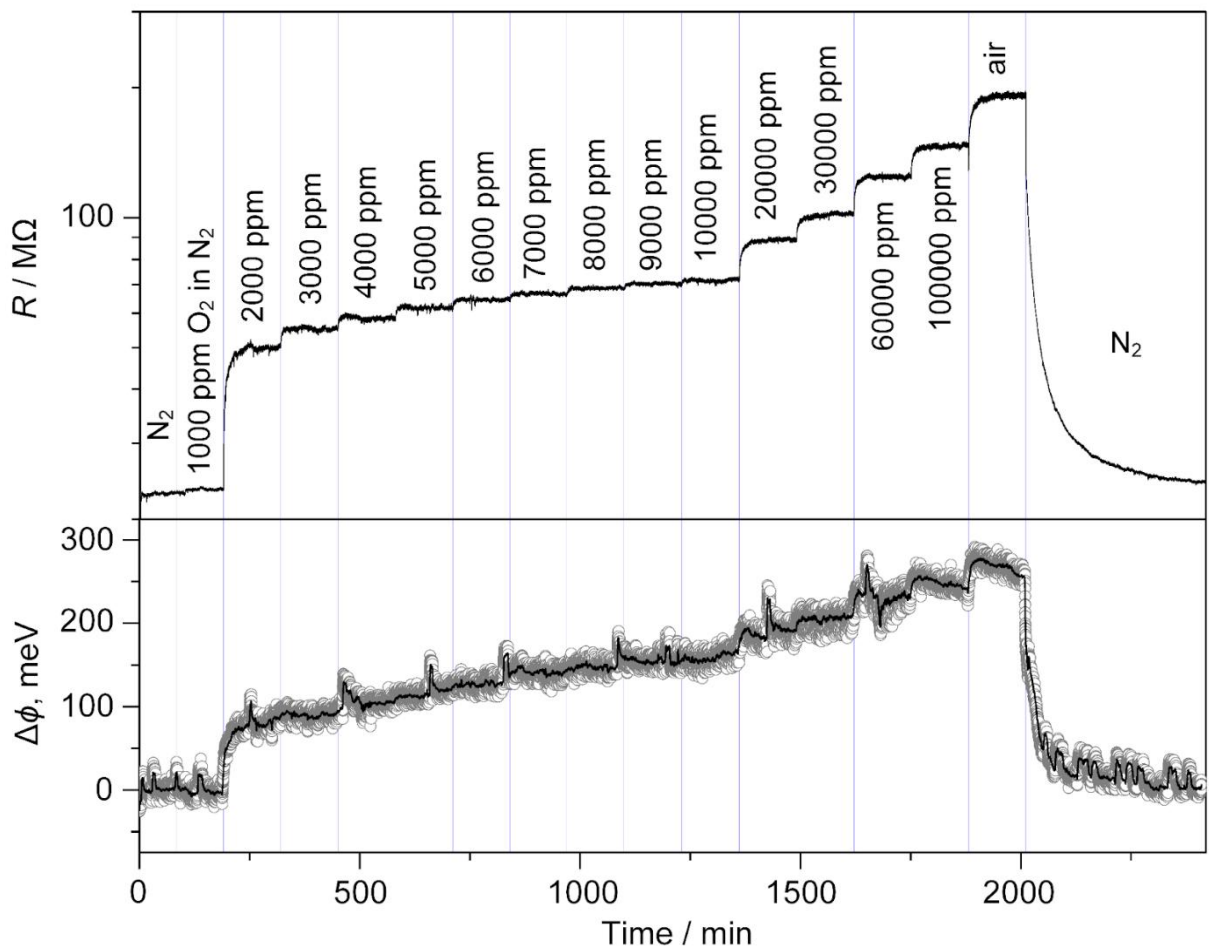

**Figure S22:** *Operando* work function measurements upon exposing $\varepsilon$-WO$_3$ to 0 – 20 vol% O$_2$ in N$_2$ at 330 °C.

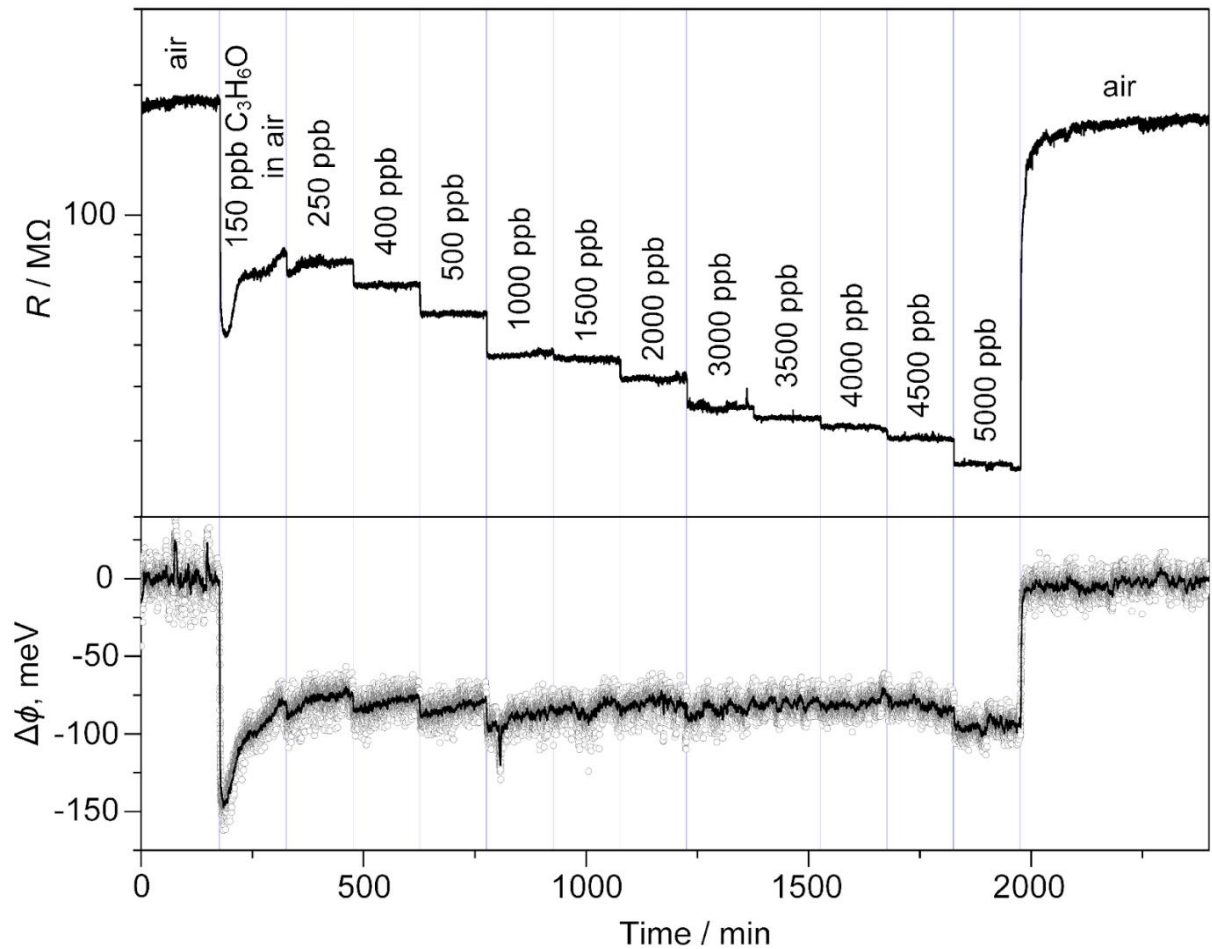

**Figure S23:** *Operando* work function measurements upon exposing *ε*-WO$_3$ to 150 – 5000 ppb C$_3$H$_6$O in air at 330 °C.

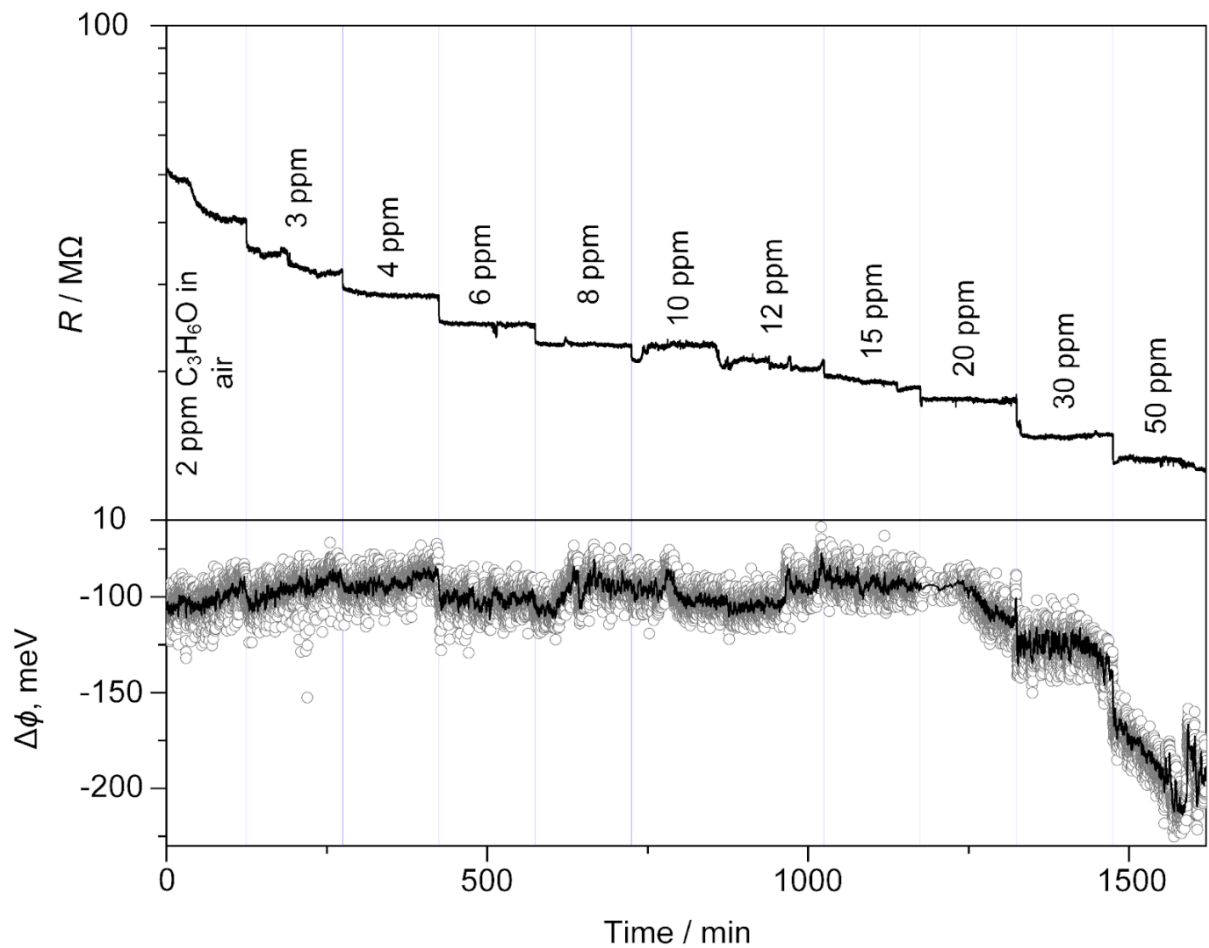

**Figure S24:** *Operando* work function measurements upon exposing $\varepsilon$-WO$_3$ to 2 – 50 ppm C$_3$H$_6$O in air at 330 °C.

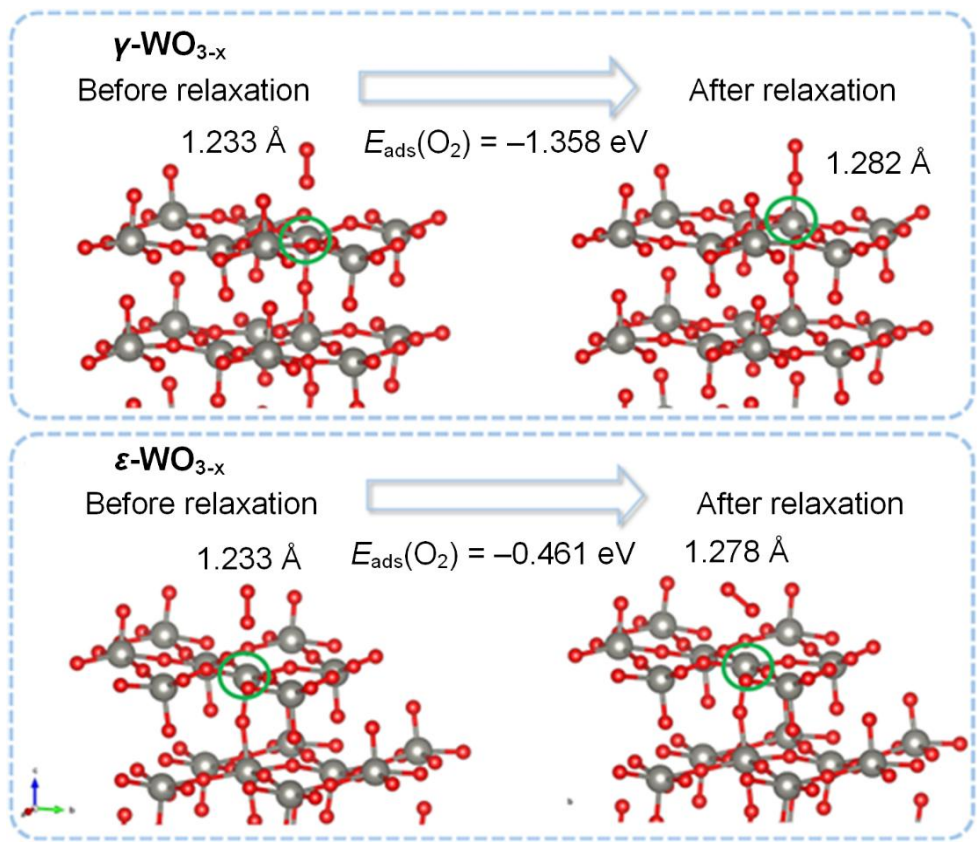

**Figure S25:** Oxygen molecule adsorbed on $\gamma$-WO$_{3-x}$ (001) and $\varepsilon$-WO$_{3-x}$ (001) at the oxygen vacancy site. Indicated are the values of $E_{ads}$ as well as O–O bond length before and after adsorption.

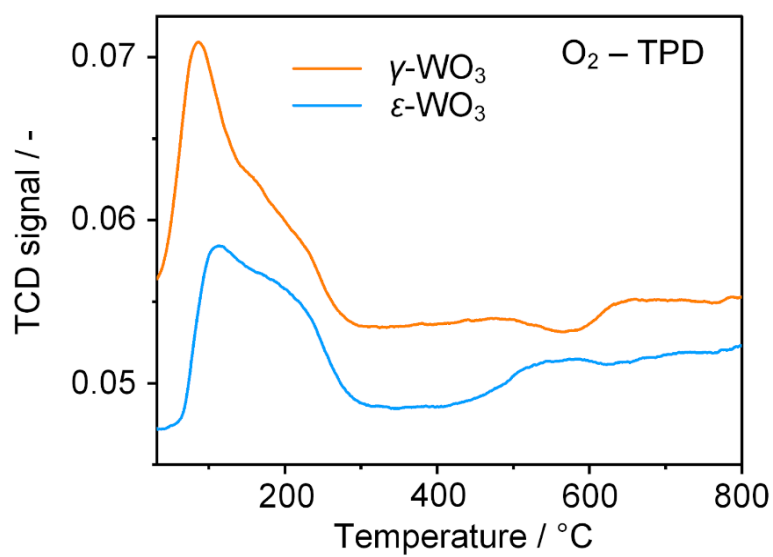

**Figure S26:** TCD signal during O$_2$-desorption from $\gamma$-WO$_3$ and $\varepsilon$-WO$_3$.

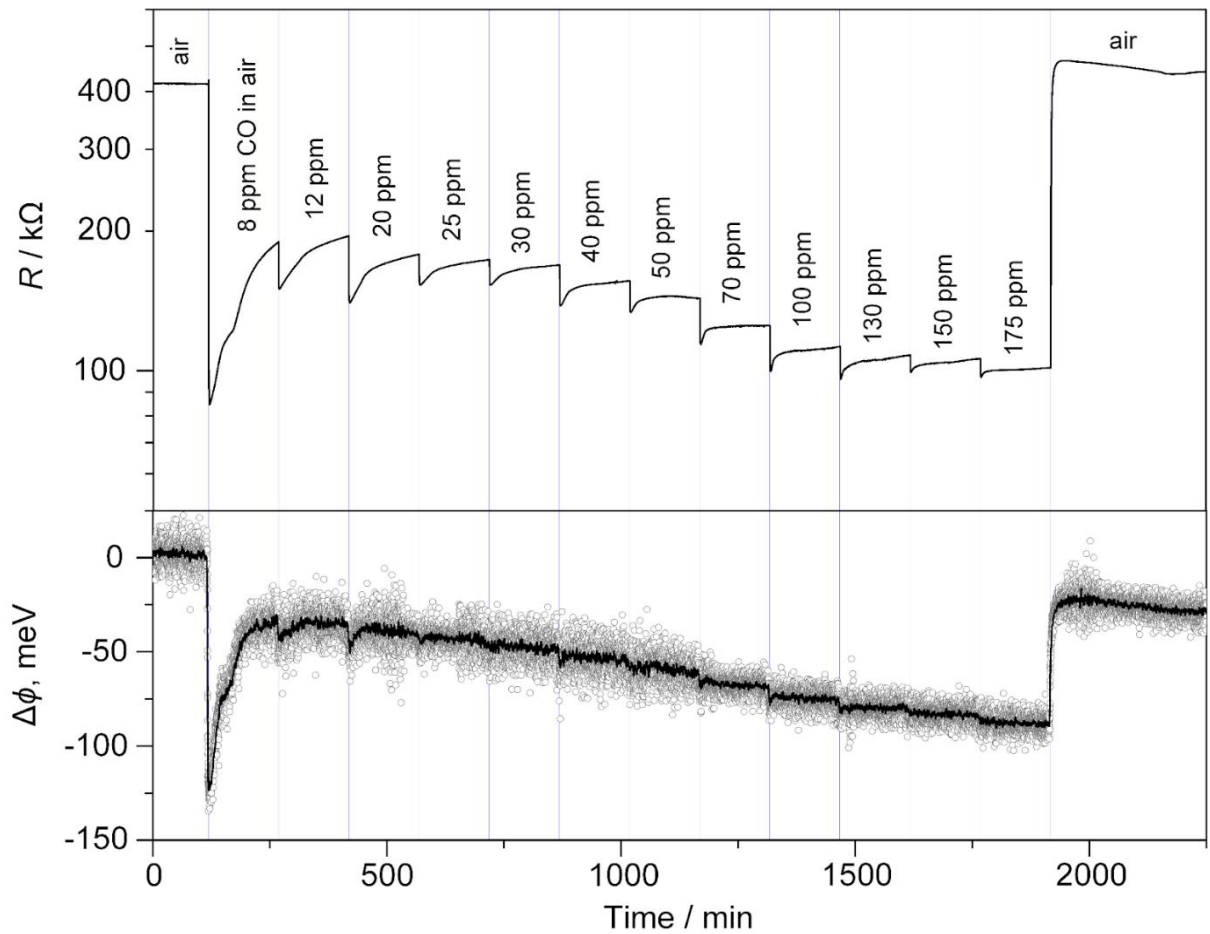

**Figure S27:** *Operando* work function measurements during exposure of SnO$_2$ to 8 – 175 ppm CO in air at 400 °C.

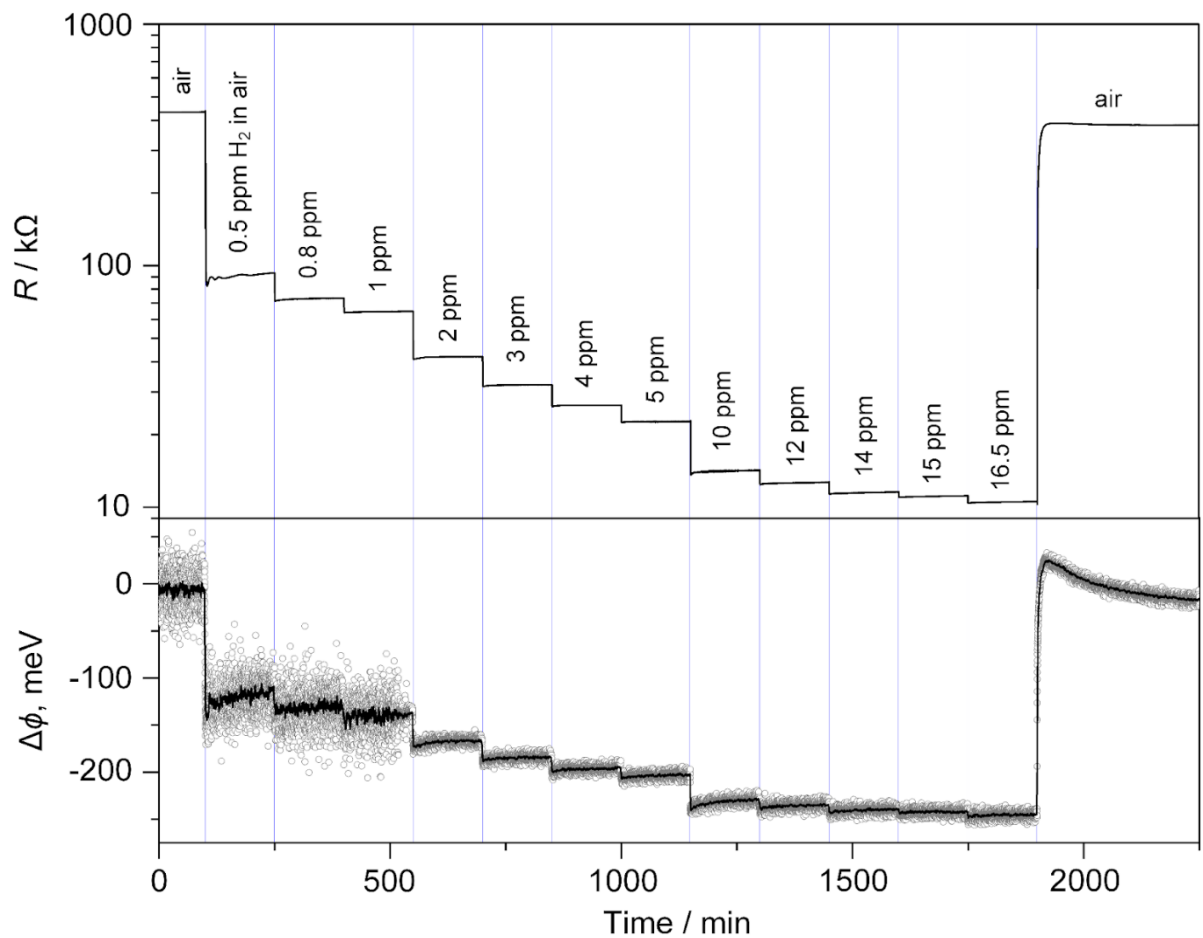

**Figure S28:** *Operando* work function measurements during exposure of $SnO_2$ to 0.5 – 16.5 ppm $H_2$ in air at 400 °C.

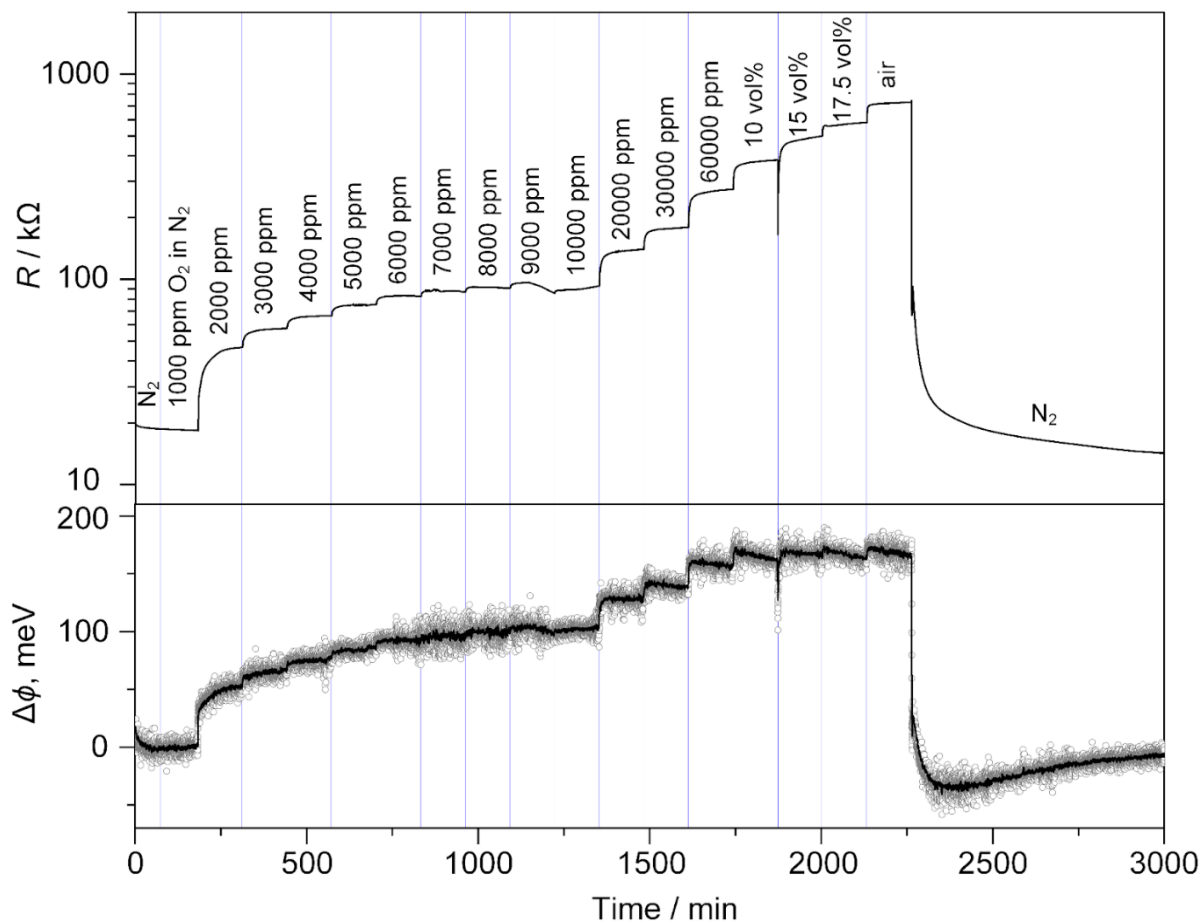

**Figure S29:** *Operando* work function measurements during exposure of SnO$_2$ to 0.1 – 20 vol% O$_2$ in N$_2$ at 300 °C.

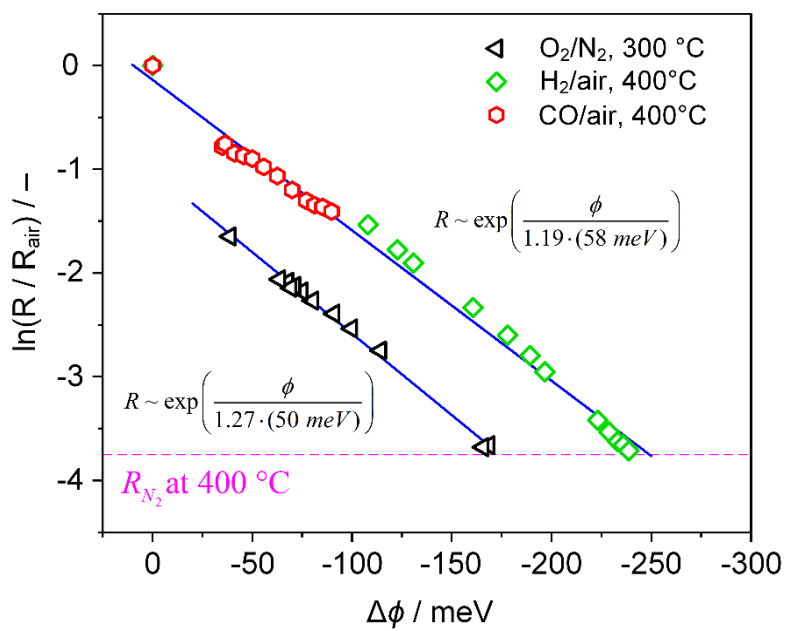

**Figure S30:** *Operando* work function results of SnO$_2$. Note that O$_2$/N$_2$ data are collected at lower temperature than O$_2$/air and H$_2$/air.

**Table S1:** XRD refinement results.

| Sample | Monoclinic $\gamma$-WO$_3$ (P2$_1$/n, #14) | | | | | | | Monoclinic $\varepsilon$-WO$_3$ (Pc, #7) | | | | | | |
|---|---|---|---|---|---|---|---|---|---|---|---|---|---|---|
| | $d_{XRD}$ / nm | $\varepsilon$ / % | $a$ / Å | $b$ / Å | $c$ / Å | $\beta$ / ° | $R_{Bragg}$ / % | $d_{XRD}$ / nm | $\varepsilon$ / % | $a$ / Å | $b$ / Å | $c$ / Å | $\beta$ / ° | $R_{Bragg}$ / % |
| [a] Pure WO$_3$ (only $\gamma$) | 23.4 | – | 7.336 | 7.511 | 7.681 | 90.780 | 1.861 | – | – | – | – | – | – | – |
| [b] Pure WO$_3$ ($\gamma$-rich, Figure S1) | 23.3 | – | 7.319 | 7.521 | 7.675 | 90.715 | 0.727 | 16.3 | – | 5.280 | 5.169 | 7.687 | 91.469 | 0.702 |
| Si-stabilized $\varepsilon$-WO$_3$ | – | – | – | – | – | – | – | 19.4 | 0.253[c] | 5.278 | 5.235 | 7.658 | 90.767 | 0.934 |

[a] Diffractogram of pure WO$_3$ analyzed with a $\gamma$-only contribution.
[b] Diffractogram of pure WO$_3$ analyzed with a mixture of $\gamma$- and $\varepsilon$-polymorphs, quantified at ~ 80 and 20 wt%, respectively.
[c] Refinement of microstrain in the other cases yielded $\varepsilon(\%) \approx 10^{-5}$, and was therefore not included in the model.

**Table S2:** Analysis of W L$_3$-edge EXAFS. The (passive) amplitude reduction factor ($S_0^2$) is determined (from W foil) to be 0.868 ± 0.075. The $E_0$ parameters are kept constant for the different scattering paths of each spectrum. The α- and β-sites refer to (multi-)scattering calculations performed considering as absorber distinct crystallographic sites of WO$_3$ (ICSD #80056), i.e., with fractional coordinates (0.2513, 0.0277, 0.2865) and (0.2481, 0.0342, 0.7815), respectively.

| | | | | |
|---|---|---|---|---|
| **W foil** *R*-factor: 0.021 $E_0$ = 5.42 ± 0.91 eV | | | | |
| Scattering path | $N$ / – | $R$ / Å | $\sigma^2$ / Å$^2$ | |
| W – W(1) | 8[a] | 2.736 ± 0.004 | 0.0028 ± 0.0004 | First shell |
| W – W(2) | 6[a] | 3.154 ± 0.007 | 0.0034 ± 0.0006 | Second shell |
| W – W(1) – W(2) | 48[a] | 4.313 ± 0.005 | 0.0045 ± 0.0005 | Multiple scattering |
| **γ-WO$_3$** R-factor: 0.019 $E_0$ = 12.59 ± 1.89 eV | | | | |
| (α) W – O(1) | 2.74 ± 0.25 | 1.788 ± 0.013 | 0.0015 ± 0.0002[b] | Crystallographic site (α) |
| (β) W – O(1) | 1.37 ± 0.12 | 1.946 ± 0.013 | 0.0015 ± 0.0002[b] | Crystallographic site (β) |
| (β) W – O(2) | 1.37 ± 0.12 | 2.438 ± 0.013 | 0.0015 ± 0.0002[b] | Crystallographic site (β) |
| **ε-WO$_3$** R-factor: 0.016 $E_0$ = 12.40 ± 1.93 eV | | | | |
| (α) W – O(1) | 2.62 ± 0.25 | 1.787 ± 0.012 | 0.0010 ± 0.0004[c] | Crystallographic site (α) |
| (β) W – O(1) | 1.31 ± 0.13 | 1.945 ± 0.012 | 0.0010 ± 0.0004[c] | Crystallographic site (β) |
| (β) W – O(2) | 1.31 ± 0.13 | 2.437 ± 0.012 | 0.0010 ± 0.0004[c] | Crystallographic site (β) |

[a] Fixed parameters.

[b,c] A single fitting parameter is used to describe the mean-square displacement ($\sigma^2$) of W-O in the 1$^{st}$ coordination shell.

**Table S3:** Comparison of band gaps of γ-WO$_3$ calculated using DFT-1/2 with experimental values and DFT calculations using different functionals from previous reports.

| Experiment / DFT | Code / potential | $a$ / Å | $b$ / Å | $c$ / Å | $\beta$ / ° | $E_g$ / eV | Ref |
|---|---|---|---|---|---|---|---|
| Experiment | —— | 7.33 | 7.56 | 7.72 | 90.5 | 2.6 – 3.2 | 48,65,68 |
| Exp & DFT | CASTEP/ LDA | 7.24 | 7.46 | 7.61 | 90.8 | 0.87 | 66 |
| Exp & DFT | CASTEP/ PBE | 7.24 | 7.46 | 7.61 | 90.8 | 0.96 | 66 |
| DFT | PW91 | 7.56 | 7.80 | 7.84 | 90.1 | 1.36 | 67 |
| DFT | HSE06 | 7.39 | 7.64 | 7.75 | 90.3 | 2.80 | 69 |
| DFT | Siesta/ DFT-1/2 | 7.46 | 7.68 | 7.88 | 90.4 | 2.48 | 26,30 |
| **This work** | VASP/ DFT-1/2 | 7.45 | 7.62 | 7.78 | 90.5 | 2.99 | |

**Table S4:** DFT-derived structures of $\gamma$-WO$_3$ and $\varepsilon$-WO$_3$.

|             | $a$ / Å | $b$ / Å | $c$ / Å | $\beta$ / ° |
|-------------|---------|---------|---------|-------------|
| $\gamma$-WO$_3$ | 7.448   | 7.621   | 7.776   | 90.451      |
| $\varepsilon$-WO$_3$ | 5.345   | 5.261   | 7.758   | 91.248      |

**Table S5:** V$_O$ formation energy for $\gamma$- and $\varepsilon$-polymorphs.

| System | V$_O$ formation energy / eV |
|--------|-----------------------------|
| $\gamma$-WO$_{3-x}$ (lattice V$_O$) | 3.733 |
| $\gamma$-WO$_{3-x}$ (surface V$_O$) | 3.068 |
| $\varepsilon$-WO$_{3-x}$ (lattice V$_O$) | 3.655 |
| $\varepsilon$-WO$_{3-x}$ (surface V$_O$) | 3.048 |

**Table S6:** Local Bader charge and electron transfer values following acetone adsorption.

| | γ-WO$_{3-x}$ + acetone | | | ε-WO$_{3-x}$ + acetone | |
|---|---|---|---|---|---|
| Atom | Bader charge ($q_i$) | $\Delta q_i$ / \|e\| | Atom | Bader charge ($q_i$) | $\Delta q_i$ / \|e\| |
| O | 6.720 | 0.720 | O | 6.680 | 0.680 |
| C$_1$ | 3.553 | − 0.447 | C$_1$ | 3.499 | − 0.501 |
| C$_2$ | 3.480 | − 0.520 | C$_2$ | 3.490 | − 0.510 |
| C$_3$ | 3.857 | − 0.143 | C$_3$ | 3.436 | − 0.564 |
| H$_1$ | 1.163 | 0.163 | H$_1$ | 1.228 | 0.228 |
| H$_2$ | 1.005 | 0.005 | H$_2$ | 1.183 | 0.183 |
| H$_3$ | 1.029 | 0.029 | H$_3$ | 1.146 | 0.147 |
| H$_4$ | 1.083 | 0.083 | H$_4$ | 1.023 | 0.023 |
| H$_5$ | 1.041 | 0.041 | H$_5$ | 1.150 | 0.150 |
| H$_6$ | 1.014 | 0.014 | H$_6$ | 1.123 | 0.123 |
| | $\Delta Q$ | − 0.055 | | $\Delta Q$ | − 0.040 |

Negative values of $q_i$ and/or $\Delta Q$ mean that the atom/molecule loses electrons.

Note: the numbering scheme of C and H atoms is as shown below, where the molecule is in the relaxed adsorption geometry from Figure 3a.

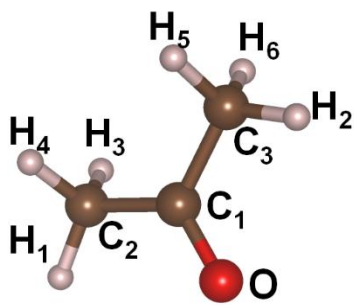

**Table S7:** Electrostatic and dipole moment ($\mu$) calculations for $\gamma$- and $\varepsilon$-WO$_3$ (001), $\gamma$- and $\varepsilon$-WO$_{3-x}$ with bulk-only V$_O$ (i.e., V$_O^{bulk}$) as well as both surface and bulk V$_O$.

| System | [a] $E_{vacuum}$ / eV | [b] $E_{Fermi}$ / eV | [c] $\phi$ / eV | $\mu$ / Debye |
|---|---|---|---|---|
| $\varepsilon$-WO$_3$ (001) | 5.284 | − 2.212 | 7.496 | 0.007 |
| $\varepsilon$-WO$_{3-x}$ (001) V$_O^{bulk}$ | 5.346 | − 1.098 | 6.444 | 0.218 |
| [d] $\varepsilon$-WO$_{3-x}$ (001) top layer | 4.419 | − 1.473 | 5.892 | − 1.226 |
| [e] $\varepsilon$-WO$_{3-x}$ (001) bottom layer | 6.228 | | 7.701 | |
| $\gamma$-WO$_3$ (001) | 5.425 | − 2.043 | 7.468 | − 0.002 |
| $\gamma$-WO$_{3-x}$ (001) V$_O^{bulk}$ | 5.495 | − 0.873 | 6.368 | 0.119 |
| $\gamma$-WO$_{3-x}$ (001) | 5.526 | − 0.804 | 6.330 | − 0.993 |

[a] $E_{vacuum}$ is determined from the electrostatic potential profiles.
[b] $E_{Fermi}$ is directly determined from the static self consistency iteration.
[c] $\phi = E_{vacuum} - E_{Fermi}$
[d,e] There is asymmetry between top and bottom facets (see Figure S18).